\documentclass[twocolumn,letterpaper]{IEEEAerospaceCLS}  

\usepackage{amsmath}
\usepackage{subfig}
\usepackage{graphicx}
\usepackage{float}
\usepackage{subcaption}
\usepackage{caption}
\usepackage{color}
\usepackage{float}
\usepackage{placeins}
\usepackage[colorlinks=true, urlcolor=blue, linkcolor=black, citecolor=black]{hyperref} 

\begin{document}
\title{Modeling and Analysis of Air-to-Ground Cellular KPIs in a 5G Testbed using Android Smartphones}

\author{%
Simran Singh\\ 
Department of ECE, North Carolina State University\\
Raleigh, NC 27606\\
ssingh28@ncsu.edu
\and 
An\i l G\"{u}rses\\ 
Department of ECE, North Carolina State University\\
Raleigh, NC 27606\\
agurses@ncsu.edu
\and 
\"{O}zg\"{u}r \"{O}zdemir\\ 
Department of ECE, North Carolina State University\\
Raleigh, NC 27606\\
oozdemi@ncsu.edu
\and 
Ram Asokan\\ 
Wireless Research Center North Carolina\\
Raleigh, NC 27606\\
asokan.ram@wrc-nc.org
\and 
Mihail L. Sichitiu\\ 
Department of ECE, North Carolina State University\\
Raleigh, NC 27606\\
mlsichit@ncsu.edu
\and 
\.{I}smail G\"{u}ven\c{c}\\
Department of ECE, North Carolina State University\\
Raleigh, NC 27606\\
iguvenc@ncsu.edu
\and 
Rudra Dutta\\
Department of CSC, North Carolina State University\\
Raleigh, NC 27606\\
rdutta@ncsu.edu
\and 
Magreth Mushi \\
Department of CSC, North Carolina State University\\
Raleigh, NC 27606\\
mjmushi@ncsu.edu
}

\maketitle

\thispagestyle{plain}
\pagestyle{plain}

\begin{abstract}
The integration of cellular communication with Unmanned Aerial Vehicles (UAVs) extends the range of command and control (C2) and payload communications of autonomous UAV applications. Accurate modeling of this air-to-ground wireless environment aids UAV mission planning. Models built on and insights obtained from real-life experiments intricately capture the variations in air-to-ground link quality with UAV position, offering more fidelity for simulations and system design than those that rely on generic theoretical models designed for ground scenarios or ray-tracing simulations. In this work, we conduct multiple aerial flights at the Aerial Experimentation and Research Platform for Advanced Wireless (AERPAW) Lake Wheeler testbed site to study the variation in key performance indicators (KPIs) of a private 4G/5G cellular base station (BS) with the UAV's altitude, distance from the BS, elevation, and azimuth relative to the BS. Variations in both 4G and 5G physical layer KPIs and application layer throughput are logged and analyzed, using two Android smartphones: a Keysight Nemo device, with enhanced KPI access, obtained through a rooted operating system (OS), and a standard smartphone running a custom application that utilizes open-source Android APIs. The BS, provided by Ericsson, consisted of two sectors, such that handover events occur during UAV flights. The observed signal strength measurements are compared to theoretical predictions from free space path loss models that incorporate the cell tower's antenna radiation patterns. Furthermore, mathematical model parameters for polynomial curve approximations are derived to fit the observed data. Light machine learning approaches, namely random forests, gradient boosting regressors and simple neural networks, are also used to model KPI behaviour as a function of UAV position relative to the BS. Model performance is evaluated using metrics such as goodness-of-fit, mean average error (MAE), and root mean square error (RMSE). The insights and models generated from real-life experiments in this study can serve as accurate and valuable tools in the design, simulation, and deployment of cellular communication-based UAV systems. 
\end{abstract}

\tableofcontents

\section{Introduction}
The use of Unmanned Aerial Vehicles (UAVs) has seen a significant increase in recent years, with applications ranging from aerial photography and surveillance\cite{11008431} to delivery services and emergency disaster response~\cite{forbesUAV5GAI}. Such applications rely on communication links between the UAV and ground control stations for real-time transmission of mission-specific and command and control communications (C2). 4G and 5G cellular technologies are a viable means of serving such communication requirements, due to the widely deployed and standardized infrastructure, coupled with mature and industry-tested secure protocol stacks \cite{uavc2survey5g}. Cellular networks enable UAVs to operate beyond the line of sight and in areas where traditional communication methods may be unavailable or unreliable. 

The success of such autonomous cellular-communication based UAV systems requires accurate modeling of the air-to-ground wireless environment. Models that accurately capture the variations in air-to-ground link quality with UAV position can aid in mission planning, route optimization, and predictable operation and control. While theoretical models and ray-tracing simulations can provide a valuable reference, insights and models derived from real-world measurements can reflect the unique scenarios that arise in air-to-ground communication links, as UAV flight trajectories extend beyond the antenna's main lobes. Besides signal strength models, cellular network key performance indicators (KPIs) such as reference signal received power (RSRP), reference signal received quality (RSRQ), channel quality index (CQI), and application layer throughput provide quantifiable and interpretable metrics for system design.

4G/LTE and 5G cellular KPIs have been analyzed, for terrestrial environments, \cite{4g_kpi_analysis},~\cite{5g_kpi_measurement}, both with public cellular networks~\cite{11008431}, and using dedicated experimental testbeds~\cite{xylouris2021experimentation}. The benefits of integrating cellular technology with UAVs~\cite{wu2021comprehensive}, has motivated such performance analysis for aerial scenarios as well~\cite{khawaja2019survey}. For example, authors in~\cite{uav_lte_performance} conducted experiments to evaluate the performance of LTE networks to support UAV applications, while \cite{uav_5g_performance} analyzed the performance of 5G networks to support UAV applications. 4G LTE RSRP and RSRQ KPIs were modeled for suburban aerial scenarios, based on empirical experiments, as a function of the two-dimensional (2D) distance of the UAV from the base station (BS) and its elevation angle in~\cite{behjati2022reliable}. Polynomial, logarithmic, and support vector regression models were employed. However, 5G signal strength was not modeled nor was the free space path loss (FSPL) model evaluated. An SDR-based channel sounder was built and utilized in~\cite{anil_ag_modeling} to measure channel impulse response and path loss, which was compared with reference statistical models. Similarly, the channel impulse response was modeled in~\cite{lyu2024fixedw} using a SDR-based channel sounding with a fixed-wing aircraft, for a rural region, using Rician fading model. However cellular KPIs such as channel rank and channel quality index (CQI) were not measured in these studies. The end to end  performance of cellular networks optimized for terrestrial environments was evaluated in~\cite{festag2021end}, in terms of meeting the throughput and latency requirements of video transmission from a UAV, and the reliability gaurantees of the control and command (C2) link. However, signal strength metrics were not modeled and physical layer KPIs were not analyzed. 


In this work, we conduct multiple aerial flights at the Aerial Experimentation and Research Platform for Advanced Wireless (AERPAW) \cite{aerpaw_website} Lake Wheeler testbed site to study the variation in KPIs of a private 5G cellular BS with the UAV's altitude, distance from the BS, elevation, and azimuth relative to the BS. Our contributions are as follows:
\begin{itemize}
    \item Experimental analysis of UAV 4G and 5G KPIs at AERPAW's testbed with a private BS.
    \item Comparison of FSPL, polynomial, and ML models in predicting $5$G RSRP as a function of the three dimensional (3D) distance of the UAV from the BS, and elevation and azimuth angle of the UAV with respect to the BS.
    \item Open-source data and Android software for reproducibility and future research.
\end{itemize}

The rest of this paper is organized as follows. The AERPAW measurement testbed and the Android measurement software is explained in Section~\ref{section:measurementSetup}. The signal strength prediction models are defined in Section~\ref{section:predictionModels}, and their performance is evaluated in Section~\ref{section:predictionResults}. 4G and 5G cellular KPIs for various UAV trajectories are analyzed in Section~\ref{section:kpiAnalysis} and Section~\ref{section:conclusion} concludes this work.



\begin{figure}
    \centering
{\includegraphics[width=0.5\textwidth]{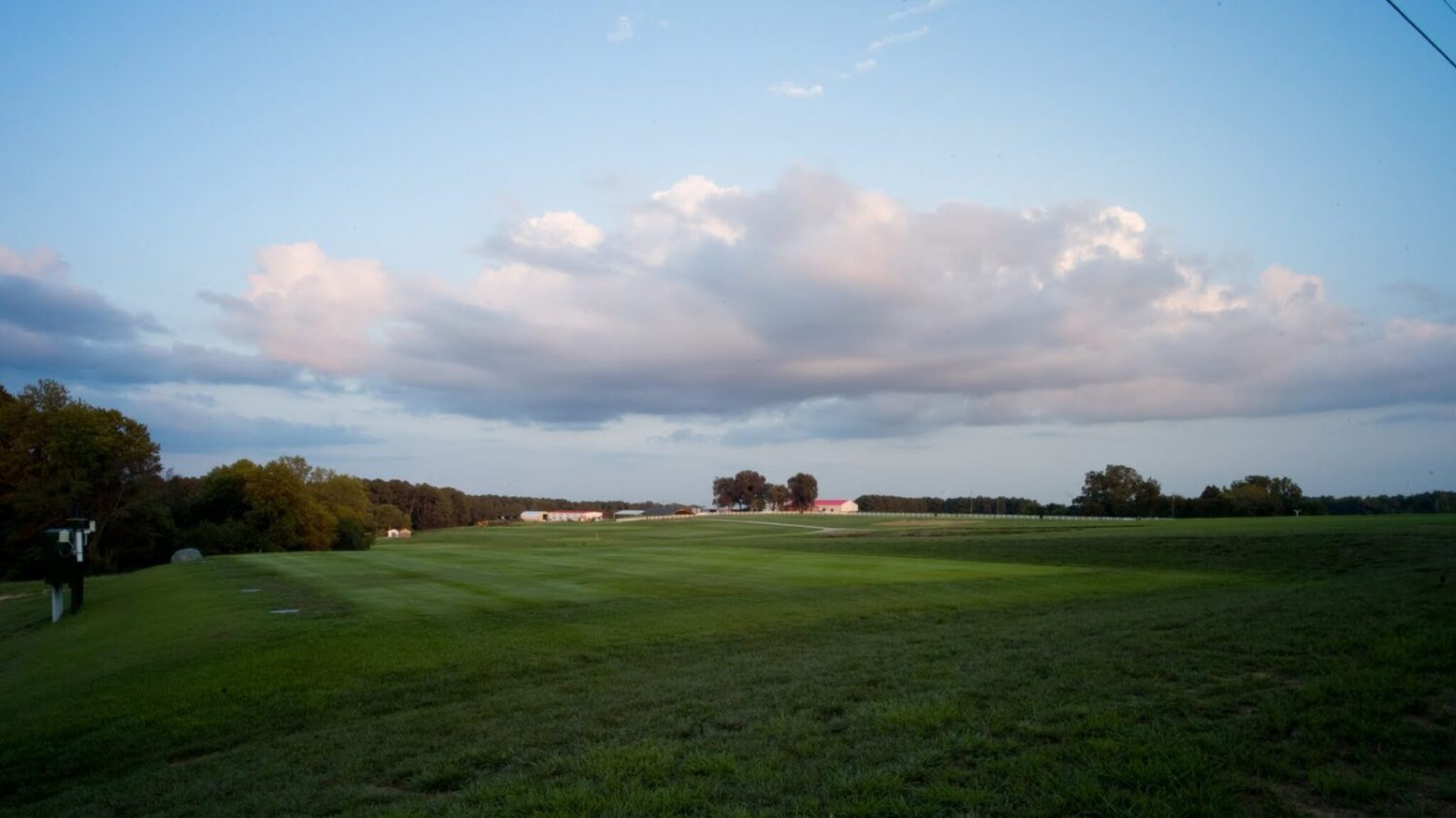}}  
    \caption{AERPAW Lake Wheeler Road Field Labs, where the UAV flights were conducted. The site has rural terrain characteristics.}
\label{fig:aerpaw_lake_wheeler_test_site}
\end{figure}

\begin{figure}[t!]
    \centering
     \subfloat[Azimuth  pattern]{\includegraphics[width=0.5\textwidth]{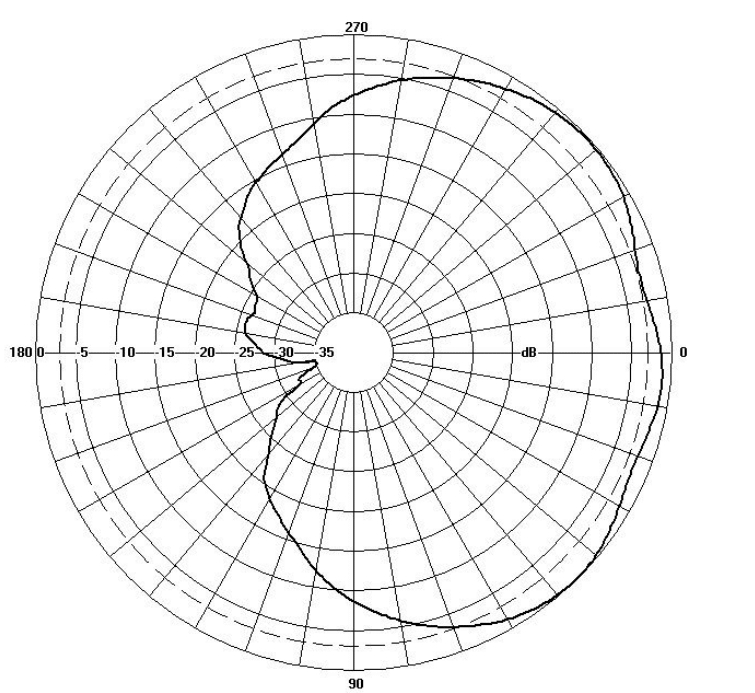}\label{fig:5G_azimuth}}
      \par
    \subfloat[Elevation  pattern] {\includegraphics[width=0.5\textwidth]{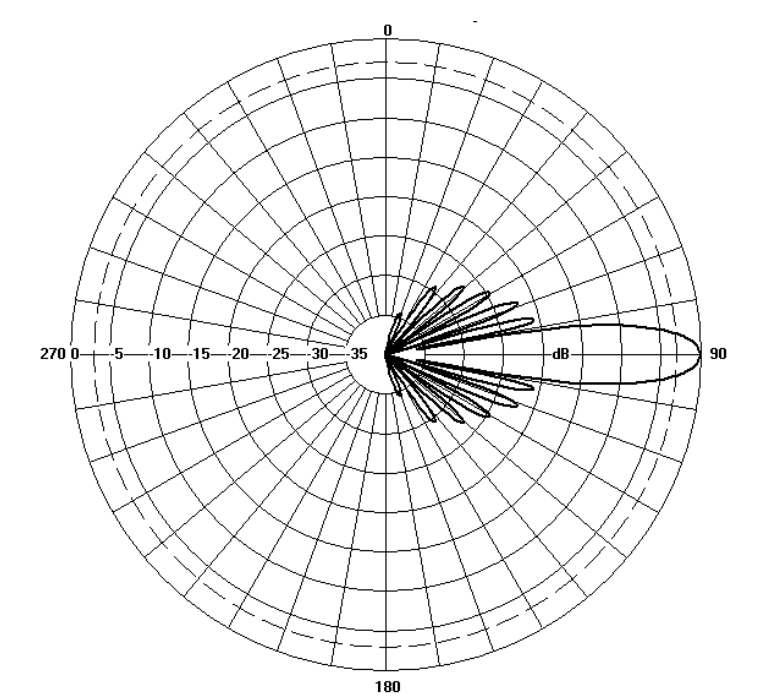}\label{fig:5G_elevation}}     
    \caption{Radiation pattern of the 5G antenna used in the AERPAW Ericsson BS, in the azimuth (a) and elevation (b) planes ~\protect\cite{alphaWireless5GAntenna}.}
\label{fig:antenna_pattern_5G}
\end{figure}

\begin{figure*}[]
\centering
\includegraphics[width=1.0\linewidth]{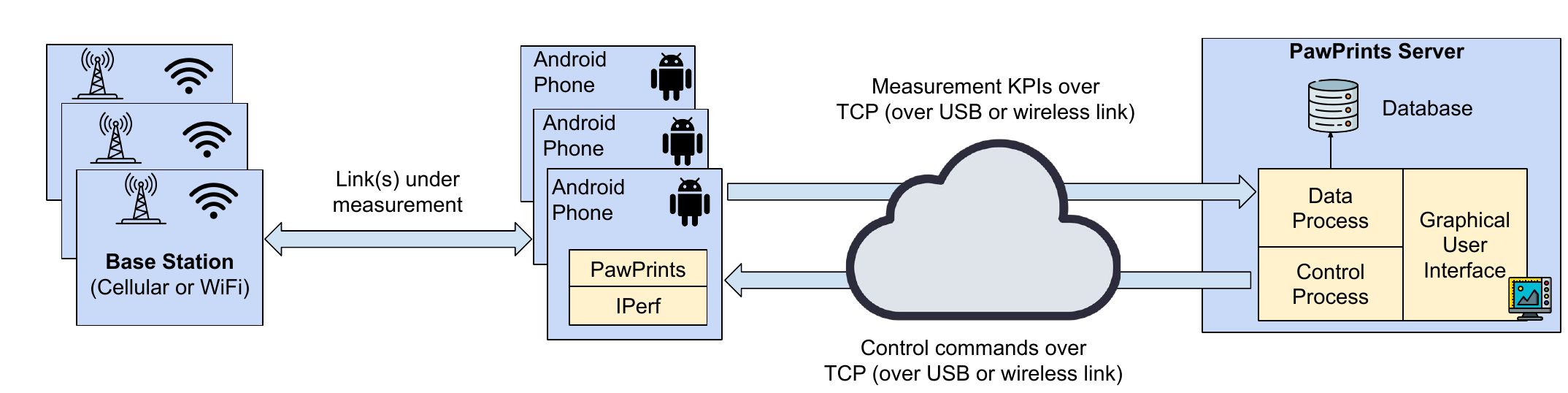}
\caption{PawPrints architecture and components~\protect\cite{pawprintsGit}.}\label{fig:pawprints_archictecture}
\end{figure*}

\section{AERPAW 4G/5G Measurement Testbed}\label{section:measurementSetup}
In this section, we provide details of the AERPAW testbed, configuration of the 4G/5G BS and antenna radiation pattern, and the Android software used to collect radio and throughput measurements.

The flights were conducted on AERPAW's Lake Wheeler Road Field Labs, which is located in Raleigh, NC and resembles a rural terrain, shown in Fig.~\ref{fig:aerpaw_lake_wheeler_test_site}. The AERPAW platform supports configurable aerial radio experiments, using custom-designed and built drones, and an open-source software control stack that supports preplanned trajectories.

The site consists of a private 5G non-standalone (5G-NSA) cellular network operating in the C-band, built using Ericsson equipment, with overlaid 5G and 4G sectors, covering the north-west direction. The $5$G new radio (NR) carrier was centered at $3.4$~GHz in band n77 with $100$~MHz of channel bandwidth, and the $4$G long term evolution (LTE) carrier was configured in band $66$~($1.7/2.1$~GHz) with $5$~MHz channel bandwidth. The 5G system supports $4\times4$ multiple input multiple output (MIMO) technologoy, while the 4G $2\times2$. Both LTE and NR carriers were set to $5$~W of transmit power per antenna port. The horizontal and vertical antenna radiation patterns of the 5G antenna are shown in Fig.~\ref{fig:antenna_pattern_5G}.

The UAV was equipped with a custom-built mount that could carry two Android phones simultaneously. The two Android devices that were used were: 1) Samsung S21 running a custom-built Android App, called PawPrints, and 2) Samsung S23+ with modified firmware and deeper access to cellular KPIs, running the Keysight Nemo Handy App~\cite{nemoHandyApp}. As shown in Fig.~\ref{fig:pawprints_archictecture}, PawPrints~\cite{pawprintsGit} consists of an Android application (app), which performs and logs the radio measurements, and an optional server, to which the Android app can be configured to stream measurements over TCP. The server can also remotely control and configure the PawPrints Android app. The app relies on Android APIs~\cite{androidTelephonyManager} to collect radio KPIs from cellular and WiFi base-stations. In addition to radio KPIs, the app can also log the phone's position (latitude, longitude, and altitude) along with the phone's speed and bearing using Android fused location APIs. 
IPerf runs as an external binary, and is executed by the PawPrints remote control script, along with parameters such as the server IP and Port, and direction of traffic (uplink or downlink).

\begin{figure}
    \centering
    \subfloat[Polygonal trajectory around the BS.]
    {\includegraphics[width=0.23\textwidth]
    {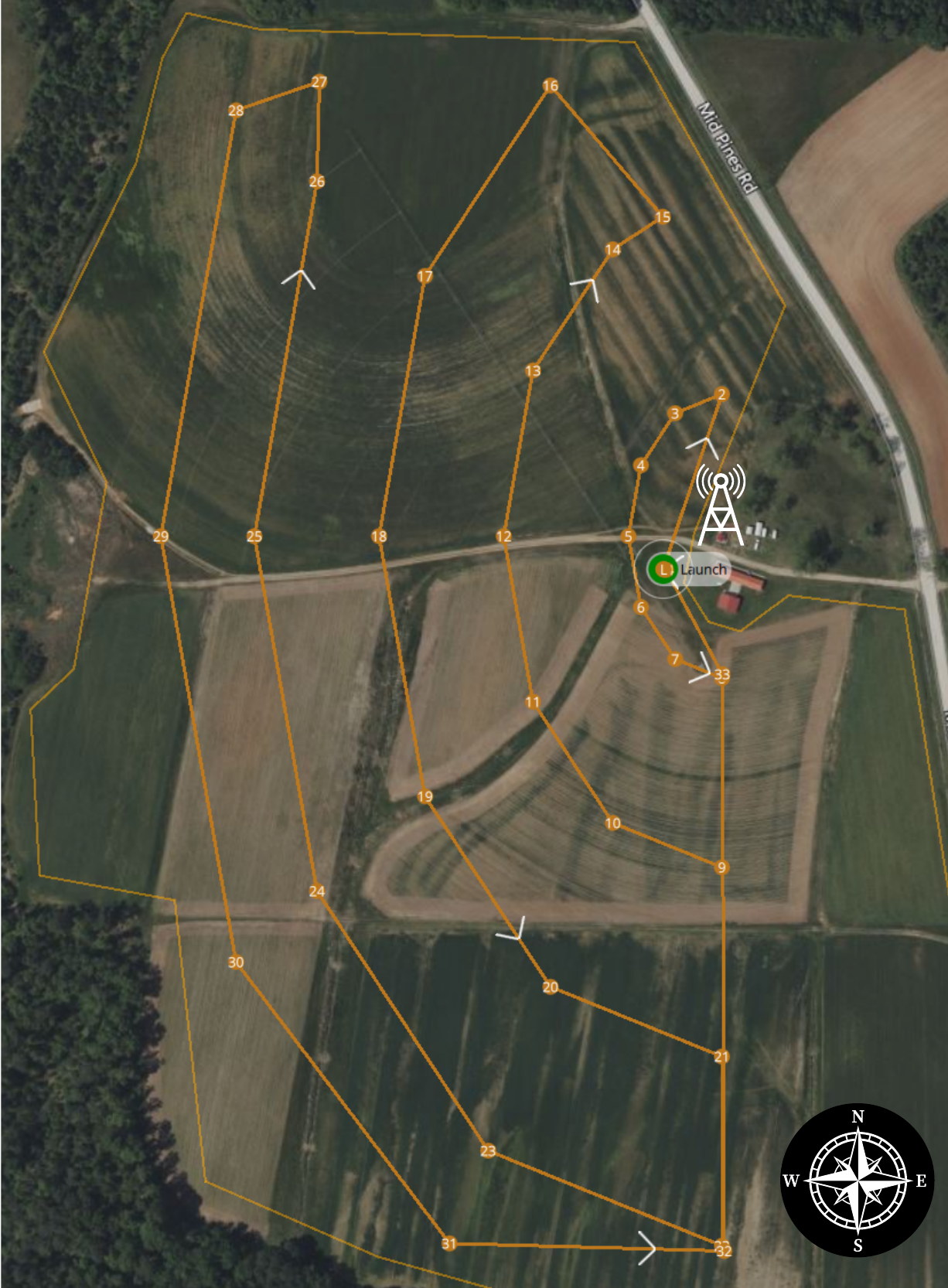}\label{fig:poly_traj}}
          \hfill
     \subfloat[Two sweeps of the UAV across the BS.]
        {\includegraphics[width=0.215\textwidth]{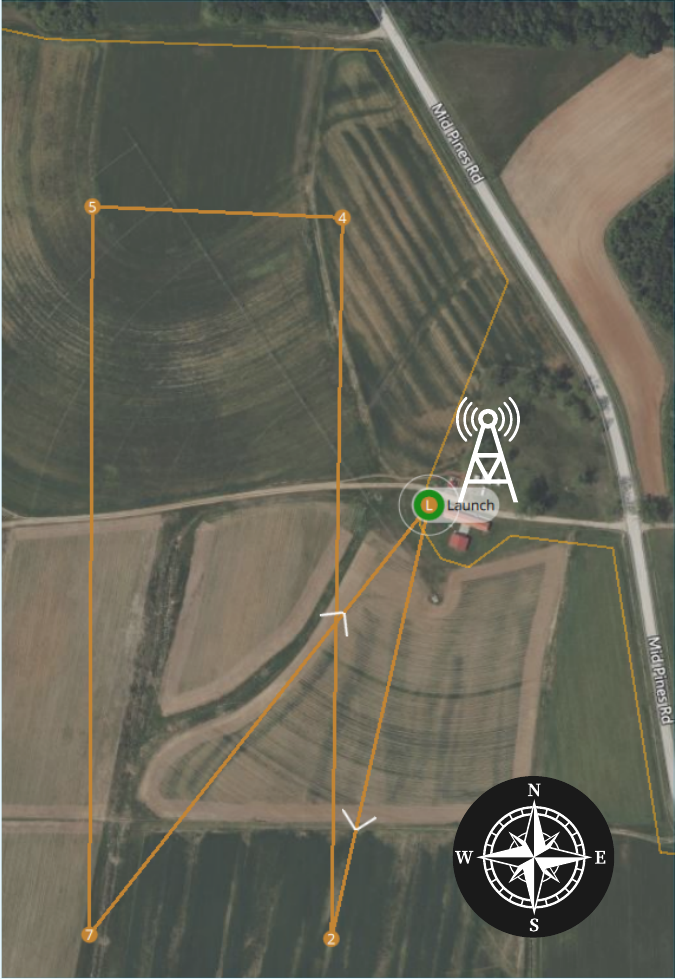}\label{fig:two_sweeps_traj}}
       \par\bigskip
       \subfloat[Horizontal sawtooth trajectory across the BS.]
        {\includegraphics[width=0.46\textwidth]{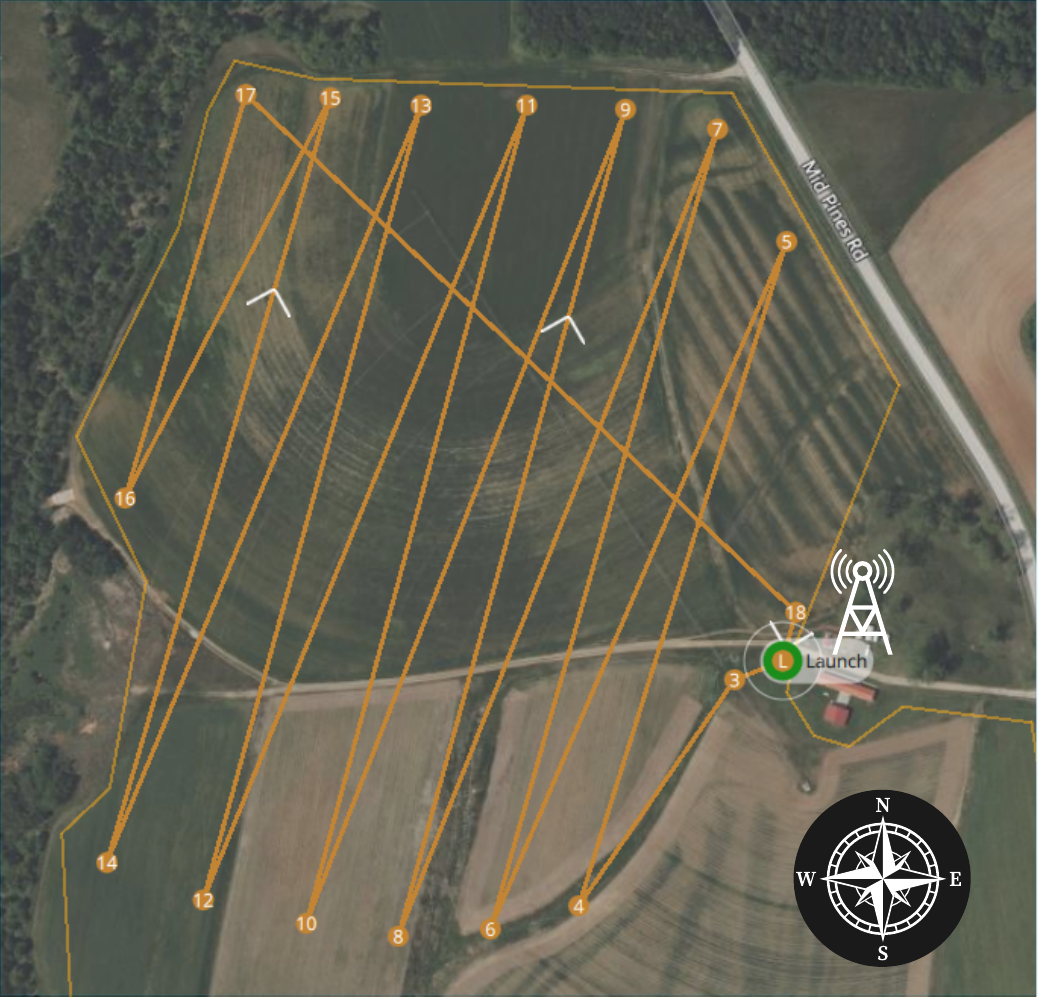}}\label{fig:sawtooth_traj}
    \caption{Flight trajectories used in the measurement campaigns. The location of the private Ericsson BS is shown as the white cell tower icon, and the UAV launch location as the green circle.}
\label{fig:measureemnt_campaigns}
\end{figure}

Three measurement campaigns were conducted, and the corresponding trajectories are shown in Fig.~\ref{fig:measureemnt_campaigns}. In the first campaign, the UAV traced a polygon around the BS and flew at an altitude of $50$~m, while logging $4$G KPIs using PawPrints. In the second campaign, the UAV traced a horizontal sawtooth trajectory, at increasing distances from the Ericsson BS. Three flights were carried out using this same pattern: two at an altitude of $30$~ m and one at an altitude of $50$~m. In the third campaign, the UAV flew across the BS, in the north-south direction, at two different distances from the BS, tracing an approximate rectangular pattern at an altitude of $30$~m. This trajectory  was flown three times. The publicly-accessible measurement datasets~\cite{aerpaw_website}
and PawPrints open source repository~\cite{pawprintsGit} enable reproduction of model training and evaluation, and support future cellular modeling research.

\section{Prediction Models for 5G SS RSRP}\label{section:predictionModels}
In this section, we define the theoretical and ML model to predict 5G signal strength.

\begin{table}[t!]
\caption{ML model parameter range for hyper-parameter tuning.}\def\arraystretch{1.7}
\label{tab:mlparameters}
\centering
\vspace{-1mm}
\footnotesize
\begin{tabular}{|p{2.3cm}|p{2.0cm}| p{3.0cm}|}
\hline \textbf {Model} & \textbf {Parameter}  & \textbf {Tuning Range}\\ \hline
Polynomial function & Degree & $2-9$ \\  \hline
Random forest, XGB & Number of decision trees & $50-300,$ in steps of $50$.  \\\hline
Random forest, XGB & Maximum tree depth & $5-20,$ in steps of $5$.  \\  \hline
Neural network & Number of hidden layers &  $[1,2]$ \\ \hline
Neural network & Neurons per layer &  $[10, 15, 20, 25, 30, 50]$ \\ \hline
Neural network & Activation function & [relu, tanh, logistic]  \\  \hline
Neural network & Solver &  [adam, lbfgs] \\  \hline
Neural network & L2 regularization rate & $[0.0001, 0.001, 0.01, 0.1]$ \\  \hline
Neural network & Initial learning rate & $[0.001, 0.01, 0.05]$  \\  \hline
\end{tabular}
\end{table}

At time instant $t$, let $\phi_{\mathrm{uav}}(t)$ denote the azimuth (horizontal) angle of the UAV with respect to the physical orientation of the BS's 5G antenna and $\theta_{\mathrm{uav}}(t)$ denote the corresponding elevation (vertical) angle. We denote the corresponding horizontal and vertical antenna gains as $G_{\mathrm{H}}(t)$ and $G_{\mathrm{V}}(t)$. We denote the total transmitted power, in watts, as $P_{\mathrm{T}}$, assumed to be constant. Then, assuming $N_{\mathrm{PRB}}$ physical resource blocks and $N_{\mathrm{SC}}$ subcarriers per resource block, the power transmitted by the BS in the synchronization signal (SS), denoted as $P_{\mathrm{SS,Tx}}$ can be calculated as:

\begin{align}
    P_{\mathrm{SS}, \mathrm{Tx}} = \frac{P_{\mathrm{T}}}{N_{\mathrm{PRB}}N_{\mathrm{SC}}}
\end{align}

Using the FSPL model, the SS RSRP at the UE at time $t$ can be calculated as:
\begin{align}
    P_{\mathrm{SS}, \mathrm{Rx}}^{\mathrm{FSPL}}(t) = 10\log_{10}(P_{\mathrm{SS,Tx}})  + G_{\mathrm{V}}(t) + \nonumber\\ G_{\mathrm{H}}(t) - 20\log_{10}(\frac{4 \pi  d_{\mathrm{uav}}(t)}{\lambda}\big),
\end{align}
where $d_{\mathrm{uav}}(t)$ represents the 3D distance of the UAV from the BS at time $t$. For simplicity, we assume the UE antenna's radiation pattern to be isotropic. 




In addition to FSPL, a multivariate polynomial regression model was used, along with the ML models built using random forest, extreme gradient boosting trees (XGB), and neural network (NN).

The polynomial model is defined as follows. Let the degree of the polynomial model be denoted as $n_{\mathrm{p}}$. Then, the polynomial function is created as follows~\cite{Lang2002Algebra}:

\begin{align}
P_{\mathrm{SS}, \mathrm{Rx}}^{\mathrm{Poly}}(t) &= \beta_{0} + \\\nonumber  &\sum_{i=0}^{n_\mathrm{p}} \; \sum_{j=0}^{n_{\mathrm{p}}-i}\;\sum_{k=0}^{n_{\mathrm{p}}-i-j} \beta_{ijk} \, (d_{\mathrm{uav}}(t))^i \, (\theta_{\mathrm{uav}}(t))^j \, (\phi_{\mathrm{uav}}(t))^k,
\end{align} where $\beta_{0}$ and $\beta_{ijk}$ represent the polynomial model parameters to be solved to minimized prediction error. Two variants of the polynomial regression model were trained: one based on logarithmic distance of the UAV from the BS and the other based on linear distance. Polynomial degrees from 2 to 9 were evaluated to analyze the trade-offs between model simplicity and closeness of fit.

Random forest and XGB are ML approaches that use decision trees as building blocks. While random forest creates decision trees independently and in parallel on random data subsets, XGB builds decision trees sequentially, where each new tree aims to reduce the errors of the previous one. Random forest and XGB ML models were trained by tuning two hyper-parameters: the number of decision trees and maximum tree depth. The performance of simple neural network (NN) models was also evaluated, where the networks were tuned by varying the number of neurons, number of hidden layers, the activation function, optimization solver, L2 regularization parameter and initial learning rate. The range of hyper-parameters is listed in Table~\ref{tab:mlparameters}. The dataset was was paritioned to two subsets based on the measurement device (S21 or S23), and separate ML and polynomials models were trained for them. A $70$-$30$ train-test ratio was utilized.

\begin{table*}[t]
\caption{Summary of SS RSRP prediction accuracy for various models.}\def\arraystretch{1.8}
\label{tab:all_model_performance}
\centering
\vspace{-1mm}
\footnotesize
\begin{tabular}{|p{0.8cm}|p{3.7cm}| p{1.5cm}|p{1.5cm}| p{1.5cm} | p{1.5cm}|} 
\hline \textbf {Device} & \textbf {Model}  & \textbf {MAE (dB)} & \textbf {RMSE (dB)} & \textbf {MAPE (\%)} & \textbf {$\mathbf{R}^{2}$}  \\ \hline
S21 & FSPL & 4.79 & 6.39 &  4.82 & 0.493 \\ \hline
S21 & Best Polynomial (degree-5) & 2.75  &  3.59 & 2.84 & 0.722\\ \hline
S21 & Stable Polynomial (degree-3)  & 2.97  & 3.79  &  3.07 &  0.691 \\ \hline
S21 & Random Forest & 2.14 & 2.89 & 2.22 &  0.82 \\ \hline
S21 & XGB & 2.26 & 2.99 & 2.31 & 0.808  \\ \hline
S21 & NN & 2.94 & 3.84 & 3.04 & 0.683  \\ \hline
S23 & FSPL & 4.25 & 5.53  & 4.32  & 0.488 \\ \hline
S23 & Best polynomial (degree-5) & 2.33  & 2.93 & 2.4 &  0.807  \\ \hline
S23 & Stable Polynomial (degree-3) & 2.62 & 3.22 & 2.69 &  0.767\\ \hline
S23 & Random Forest & 1.71 & 2.23 & 1.76 & 0.883 \\ \hline
S23 & XGB & 1.71 & 2.46 & 1.76 & 0.864  \\ \hline
S23 & NN & 3.21 & 2.59 & 2.65 & 0.769  \\ \hline
\end{tabular}
\end{table*}

\section{5G SS RSRP Modeling Results}\label{section:predictionResults}
In this section, we present 5G SS RSRP modeling results using FSPL, polynomial, and ML models, and compare performance in terms of prediction error. 

The performance of the polynomial regression model was evaluated, in terms of root mean square error (RMSE) and coefficient of determination, denoted as $\mathrm{R}^2$ score, for degrees ranging from $2-9$, and with either linear or log distance as the independent variable. Log-distance based polynomial functions yield minimum RMSE with degree $5$ polynomial functions, for S21 and S23, achieving errors of $3.59$~dB and $2.93$~dB respectively. When using linear-distance for S21, we obtain RMSE $>3.59$~dB at all polynomial function degrees, except at degree-$9$. Since higher degree polynomials tend to generalize poorly and show unpredictable behaviour, we choose the log-distance polynomial functions to fit SS RSRP, since it achieves similar or lower RMSE values at lower degrees, compared to the linear-distance variant.

\begin{figure}[]
    \centering  \subfloat{\includegraphics[width=0.5\textwidth]{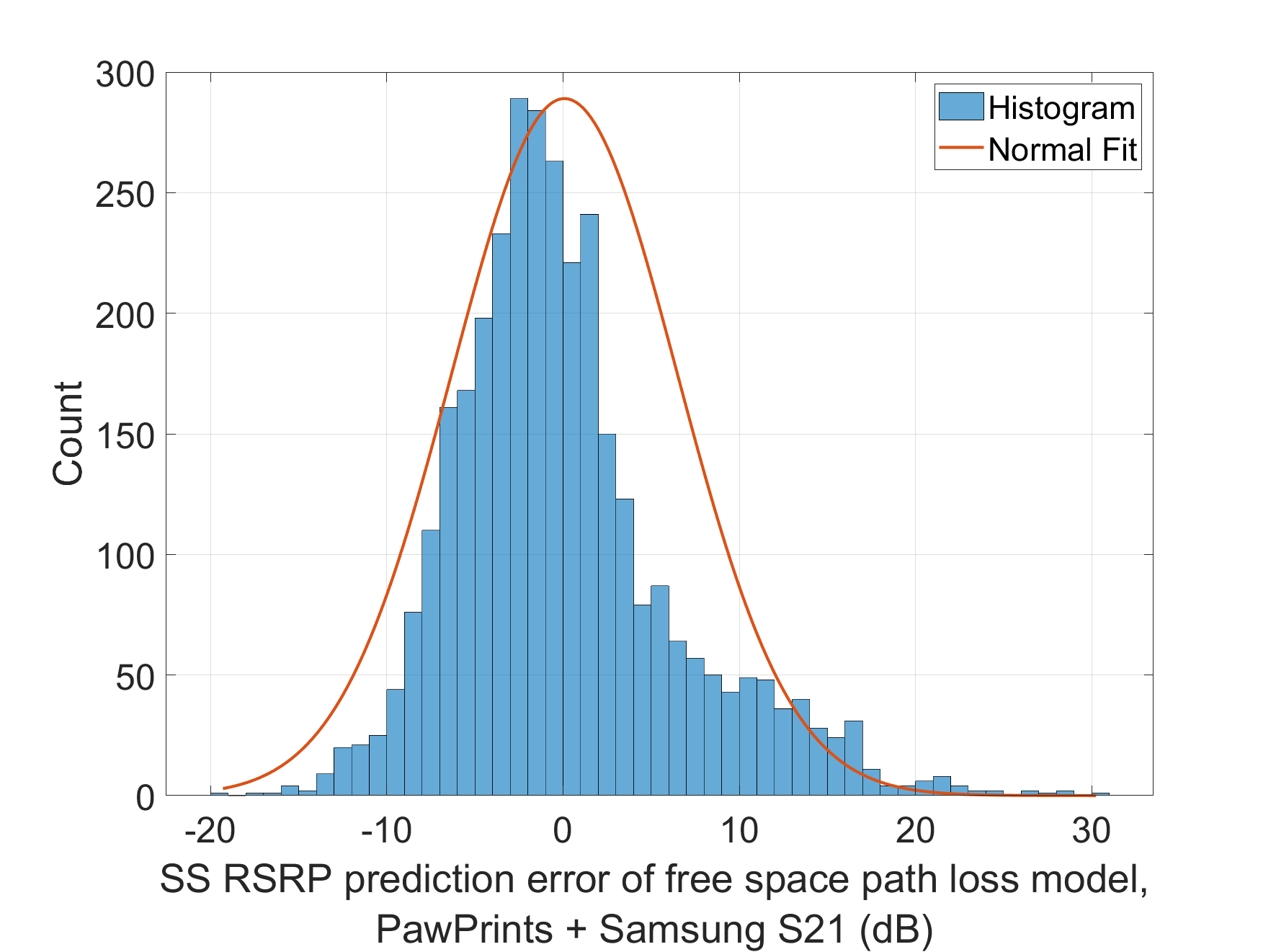}\label{fig:hist_pawprint_fspl_errors}}
      \par \bigskip
    \subfloat {\includegraphics[width=0.5\textwidth]{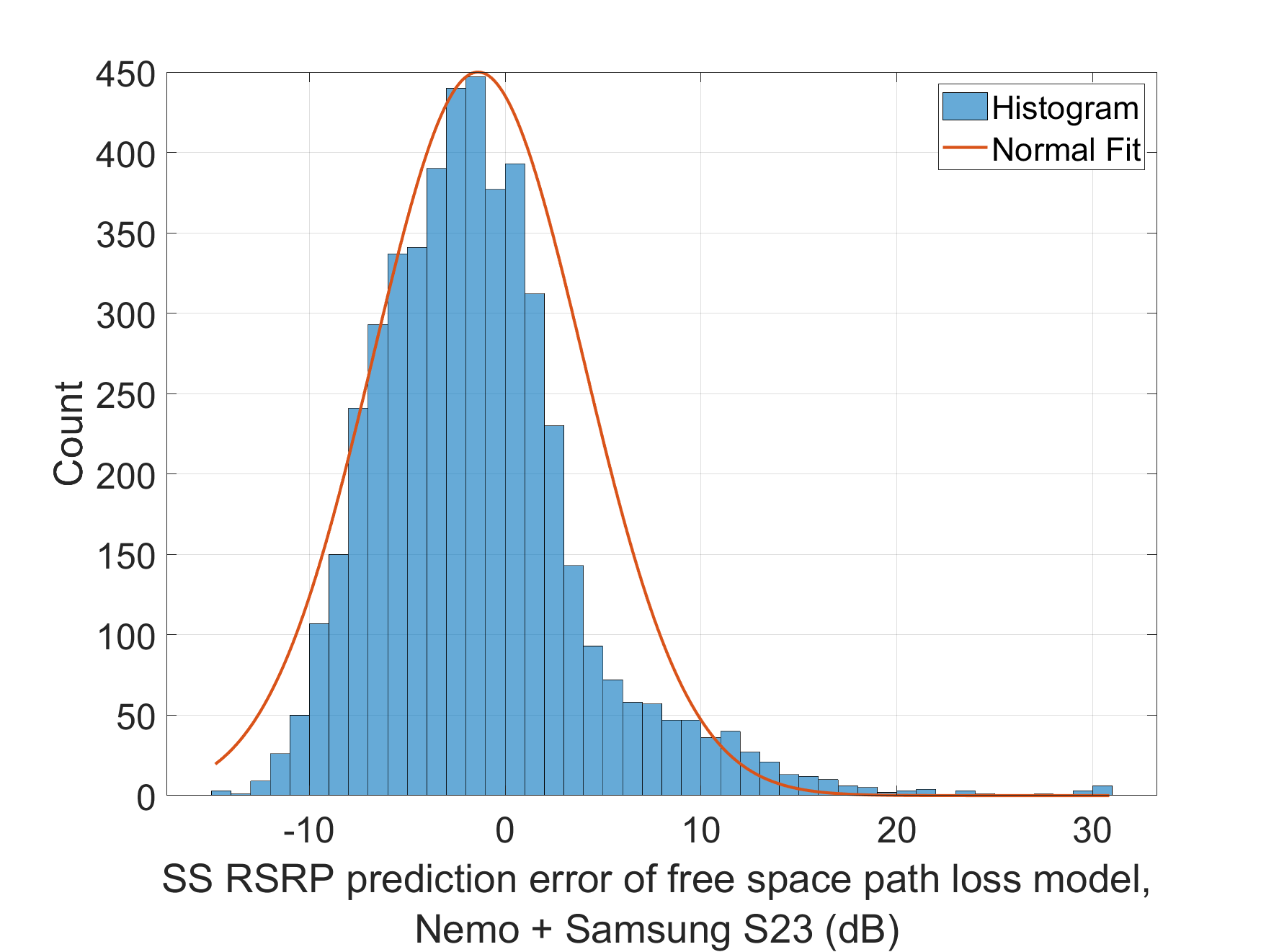}\label{fig:hist_nemo_fspl_errors}}   
    \caption{Histogram of the 5G RSRP prediction error of the free space path loss model, along with the normal distribution fit, for the Samsung S21 device in (a) and the Samsung S23 device in (b).}
\label{fig:hist_fspl_errors}
\end{figure}

The performance of all models in predicting SS RSRP is summarized in Table~\ref{tab:all_model_performance}, separately for the Samsung S21 and S23 devices. The mean absolute error (MAE), mean absolute percentage error (MAPE), root mean square error (RMSE), and $\mathrm{R}^2$ scores are computed for all models using the testing dataset. Prediction errors are defined as the difference between the observed and predicted RSRP values. Overall, the machine learning and polynomial models outperform the free-space path loss (FSPL) model. Among these, the random forest models yield the lowest prediction errors and the highest $\mathrm{R}^2$ scores. The distributions of signed FSPL-based 5G RSRP prediction errors for the Samsung S21 and S23 devices are shown in Fig.~\ref{fig:hist_pawprint_fspl_errors} and Fig.~\ref{fig:hist_nemo_fspl_errors}, respectively. The mean prediction error for S21 is $0.08$~dB with a standard deviation of $6.39$~dB and, correspondingly for S23, $-1.38$~dB with a standard deviation of $5.36$~dB.
 
\begin{figure}
    \centering
{\includegraphics[width=0.5\textwidth]{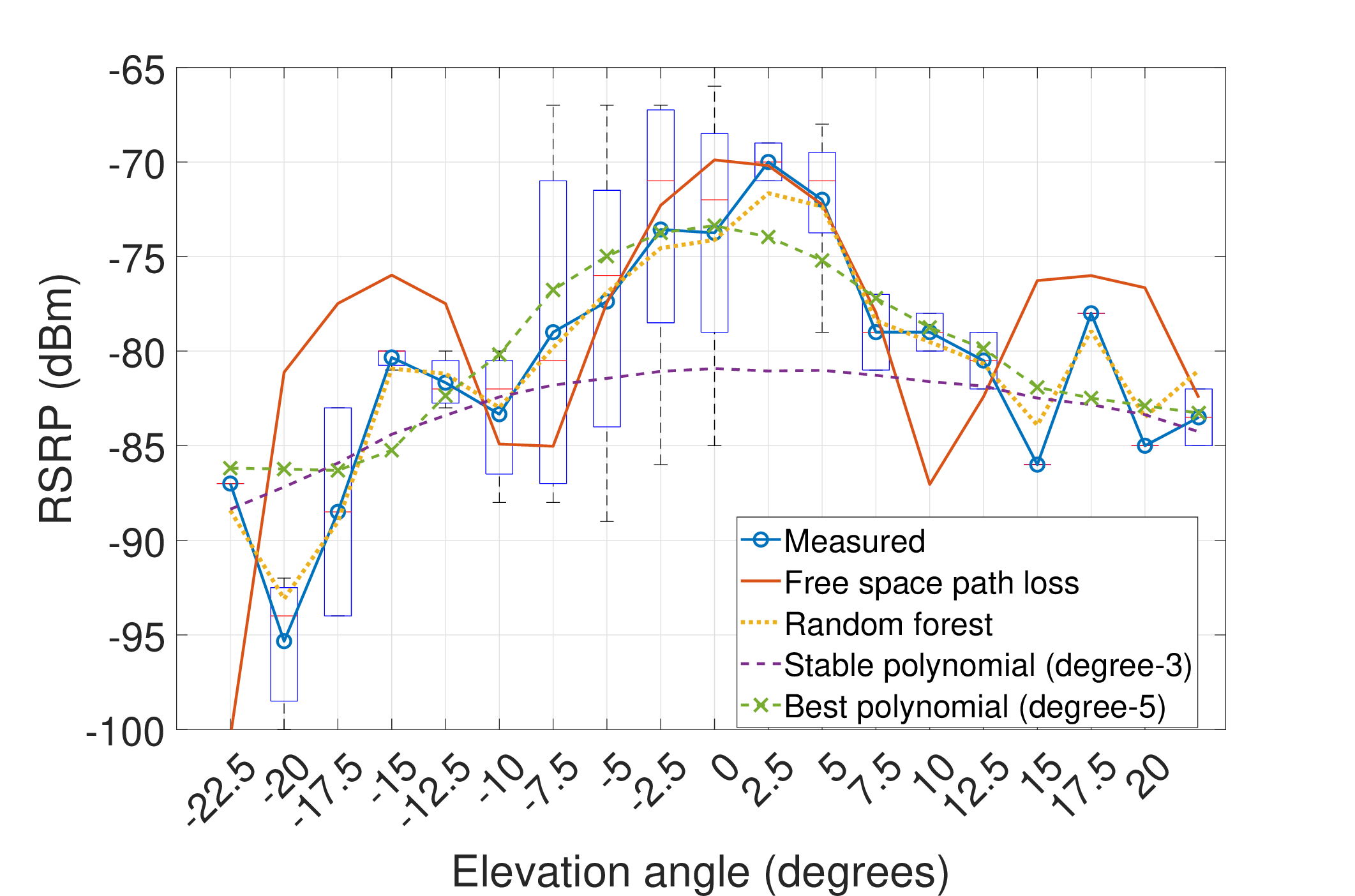}} 
    \caption{Measured and predicted RSRP during landing and takeoff, as a function of the UAV's elevation angle.}
\label{fig:landing_takeoff_rsrp}
\end{figure}

The variation in measured and predicted RSRP with elevation angle, calculated during landing and take-off, is shown in Fig.~\ref{fig:landing_takeoff_rsrp}, illustrating that the random forest and degree-$5$ polynomial models closely learn the observed measurements. Box plots of the measured RSRP are also shown, with median as the red line within the box, calculated over bins of $2.5$~degrees, The FSPL model shows deviations at 
$7.5-12.5$~degrees elevation, while the stable degree-$3$ polynomial represents a smooth approximation.

\begin{figure}[]
    \centering  \subfloat{\includegraphics[width=0.5\textwidth]{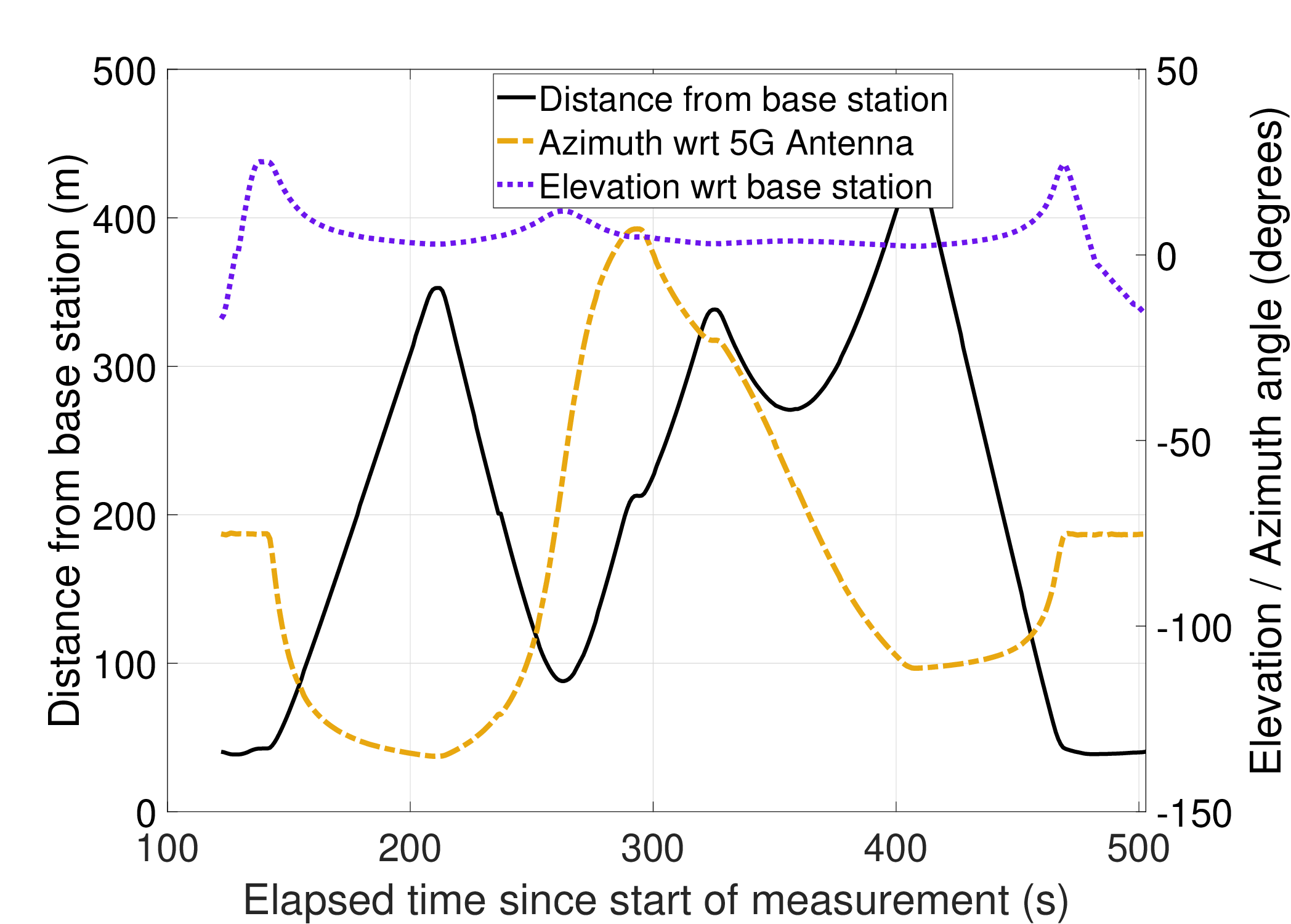}\label{fig:two_sweeps_trajectory}}
      \par \bigskip
    \subfloat {\includegraphics[width=0.5\textwidth]{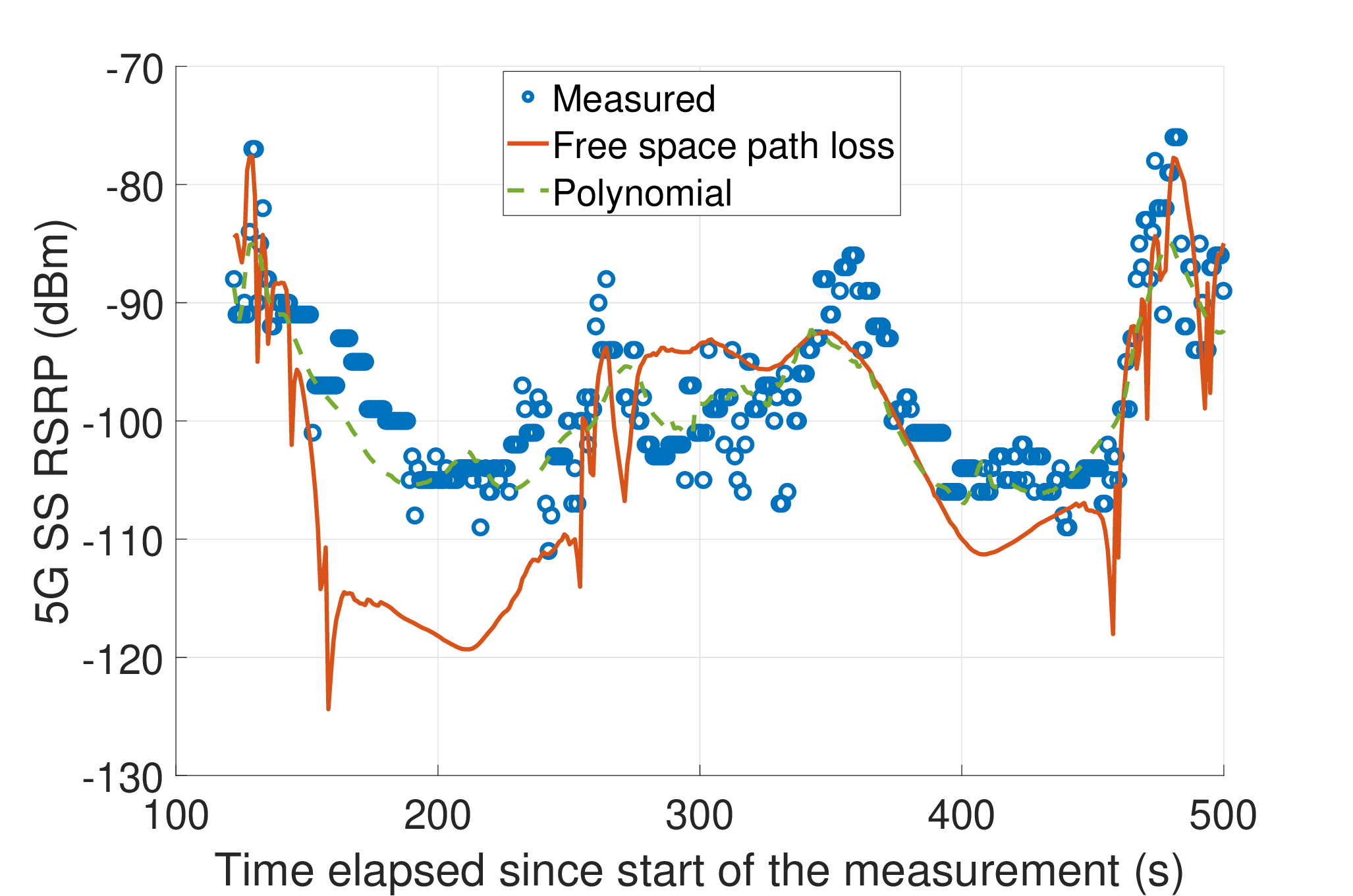}\label{fig:two_sweeps_rsrp}}     
    \caption{UAV position and orientation (a), and free-space 5G SS RSRP modelling results (b), for the two-sweeps trajectory.}
\label{fig:two_sweeps_details}
\end{figure}

\begin{figure}[]
    \centering
     \subfloat{\includegraphics[width=0.5\textwidth]{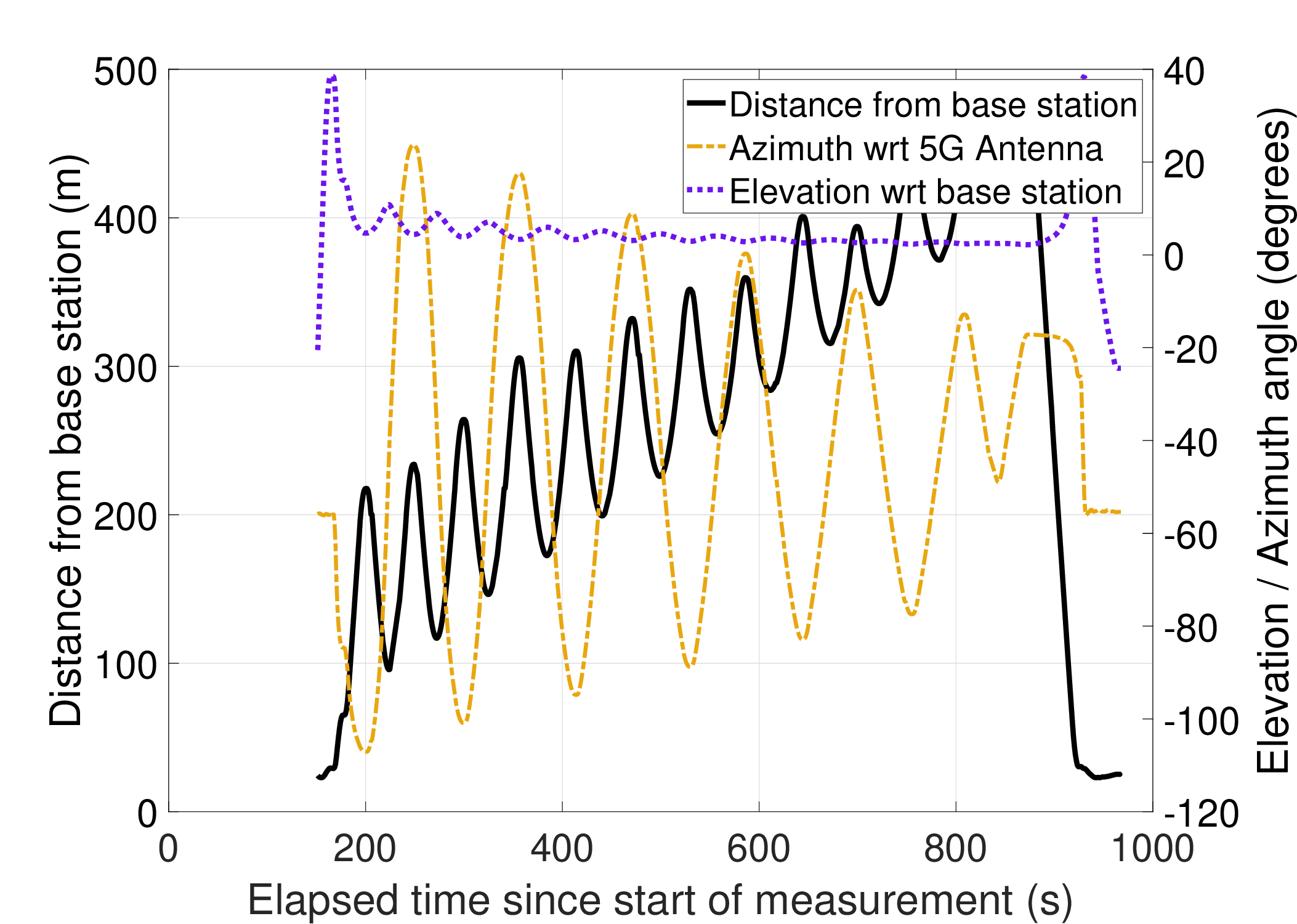}\label{fig:horizontal_sweeps_30m_trajectory}}
     \par \bigskip
    \subfloat {\includegraphics[width=0.5\textwidth]{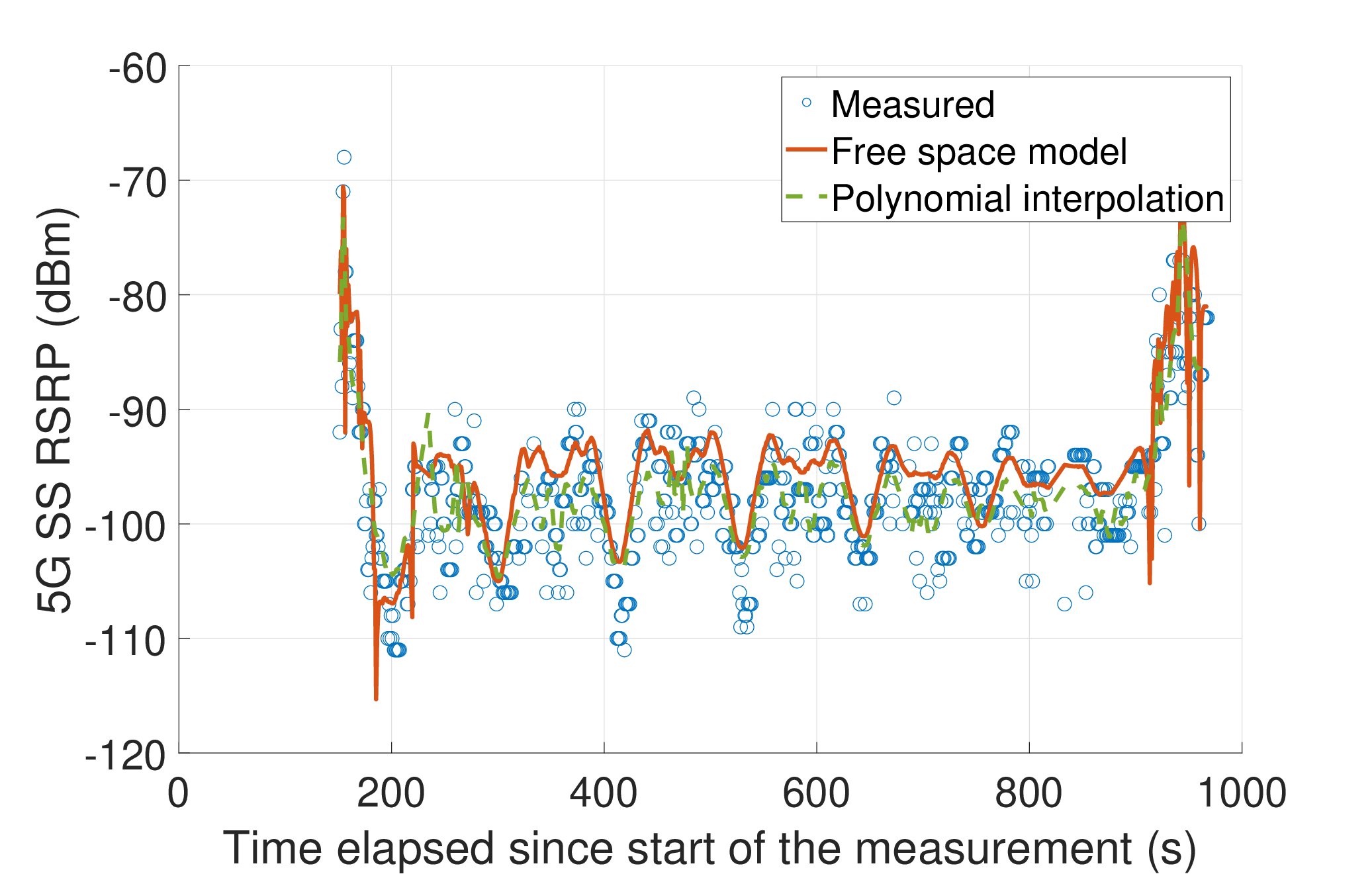}\label{fig:horizontal_sweeps_30m_rsrp}}     
    \caption{UAV position and orientation (a) and free-space 5G SS RSRP modeling results (b), for horizontal sweeps trajectory at $30$~m altitude.}
\label{fig:horizontal_sweeps_30m_details}
\end{figure}

\begin{figure}[]
    \centering
    \subfloat{\includegraphics[width=0.47\textwidth]{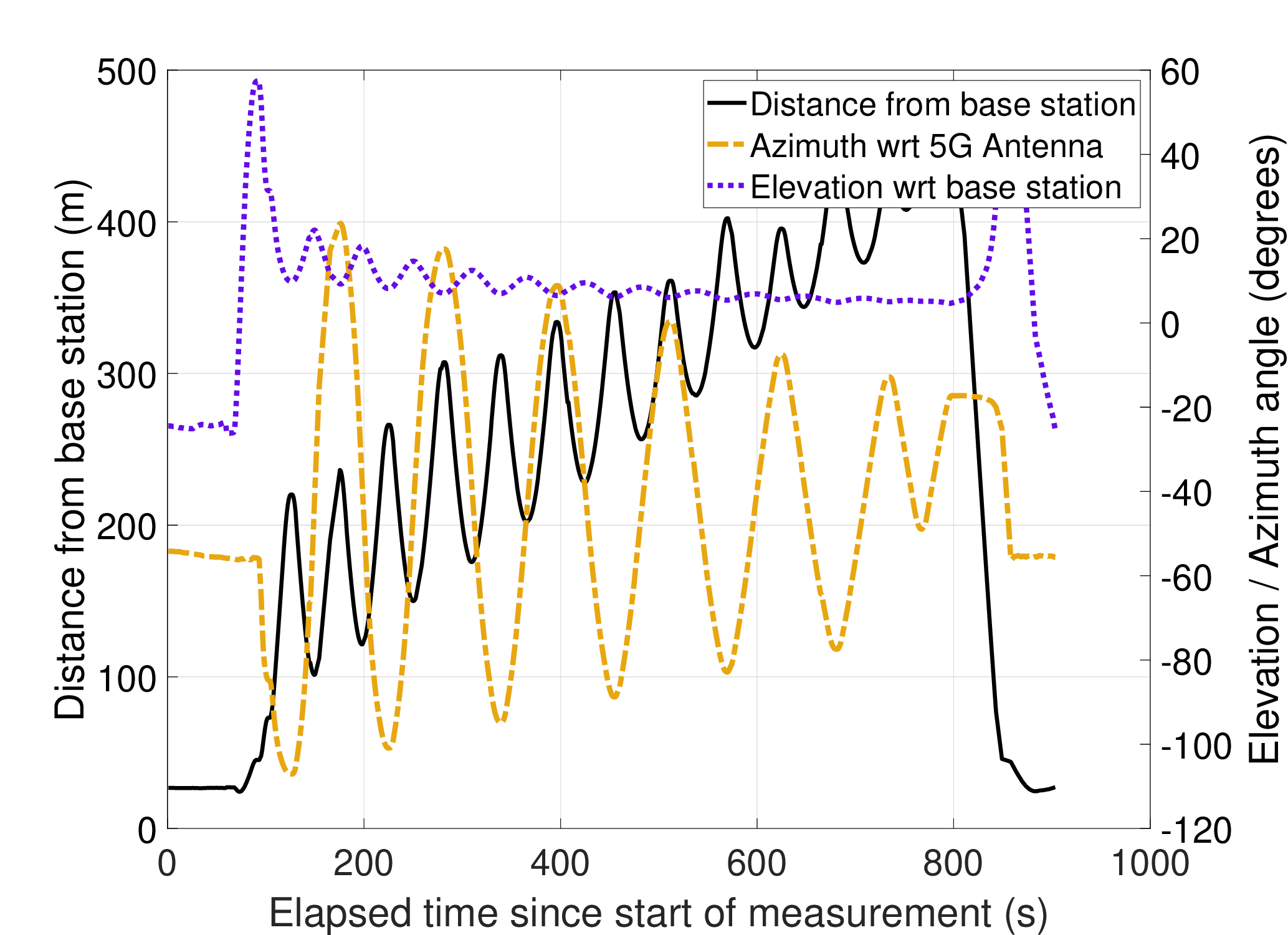}\label{fig:horizontal_sweeps_50m_trajectory}}
      \par \bigskip
    \subfloat {\includegraphics[width=0.47\textwidth]{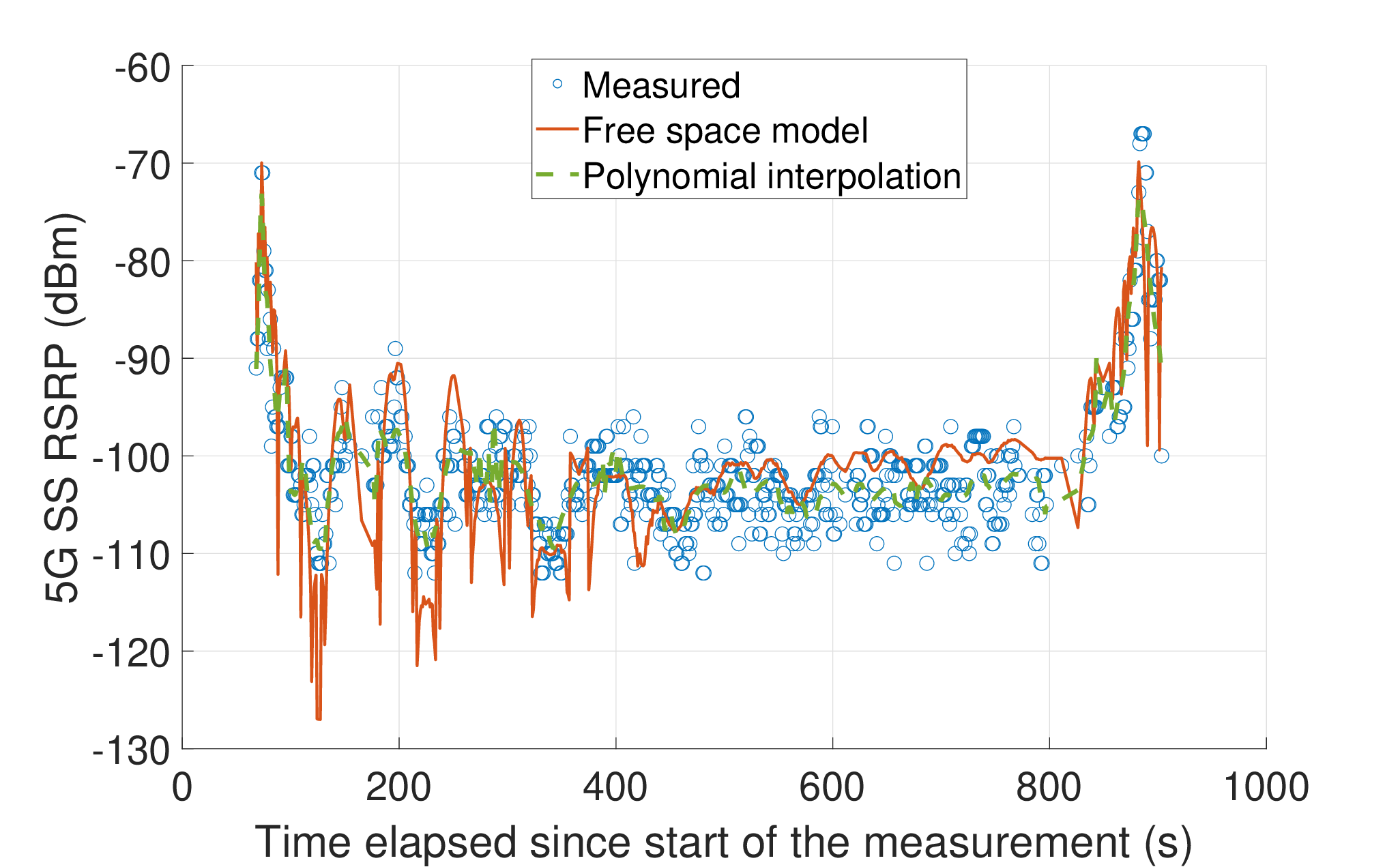}\label{fig:horizontal_sweeps_50m_rsrp}}     
    \caption{UAV position and orientation (a) and free-space 5G SS RSRP modeling results (b), for horizontal sweeps trajectory at $50$~m altitude.}
\label{fig:horizontal_sweeps_50m_details}
\end{figure}

For a detailed view into the signal strength behaviour, the variation in measured and calculated 5G SS RSRP with time was plotted, along with the UAV's position at each time instant.Fig.~\ref{fig:two_sweeps_details} shows these plots for the two-sweeps trajectory,  Fig.~\ref{fig:horizontal_sweeps_30m_details} for the horizontal sawtooth trajectory at an altitude of $30$~m, and Fig.~\ref{fig:horizontal_sweeps_50m_details} for the horizontal sawtooth trajectory at an altitude of $50$~m, all for the S23 device. The distance of the UAV from the BS is shown as the black plot, azimuth angle with respect to the 5G antenna as the dash-dot yellow plot, and the elevation angle of the UAV with respect to the 5G antenna as the dotted purple plot, in Fig.~\ref{fig:two_sweeps_trajectory}, Fig.~\ref{fig:horizontal_sweeps_30m_trajectory}, and Fig.~\ref{fig:horizontal_sweeps_50m_trajectory}, as a function of elapsed time. Measured 5G RSRP values are shown as the hollow blue circles, free-space path loss model RSRP values as the red plot, and the polynomial function's predictions as the dash-dot green plot in Fig.~\ref{fig:two_sweeps_rsrp}, Fig.~\ref{fig:horizontal_sweeps_30m_rsrp}, and Fig.~\ref{fig:horizontal_sweeps_50m_rsrp}. 

From Fig.~\ref{fig:two_sweeps_trajectory} and Fig.~\ref{fig:two_sweeps_rsrp}, the large error in the FSPL model's predictions can be seen at the time instants where the azimuth angle is outside the main lobe. The mean FSPL prediction error over this flight is relatively high, $7.24$~dB, with a standard deviation of $7.85$~dB. Similarly, Fig.~\ref{fig:horizontal_sweeps_50m_trajectory} and Fig.~\ref{fig:horizontal_sweeps_50m_rsrp} illustrate a scenario where the error in FSPL model is large due to large variations in the elevation angle of the UAV, beyond $40^\circ$. In this scenario, the MAE of FSPL is $4.27$~dB with a standard deviation of $5.66$~dB. The performance of the FSPL model is comparatively better over the horizontal sawtooth trajectory at $30$~m altitude, where the variation in elevation and azimuth angles is lower. The MAE over this flight is $3.49$~dB and the standard deviation of absolute error is $3.98$~dB. The absence of any extreme errors for this measurement flight can be visually confirmed from Fig.~\ref{fig:horizontal_sweeps_30m_trajectory} and Fig.~\ref{fig:horizontal_sweeps_30m_rsrp}.
\section{5G Cellular KPI Analysis}\label{section:kpiAnalysis}
\begin{figure*}[]
    \centering
     \subfloat[LTE RSRP (dBm)] {\includegraphics[width=0.33\textwidth]{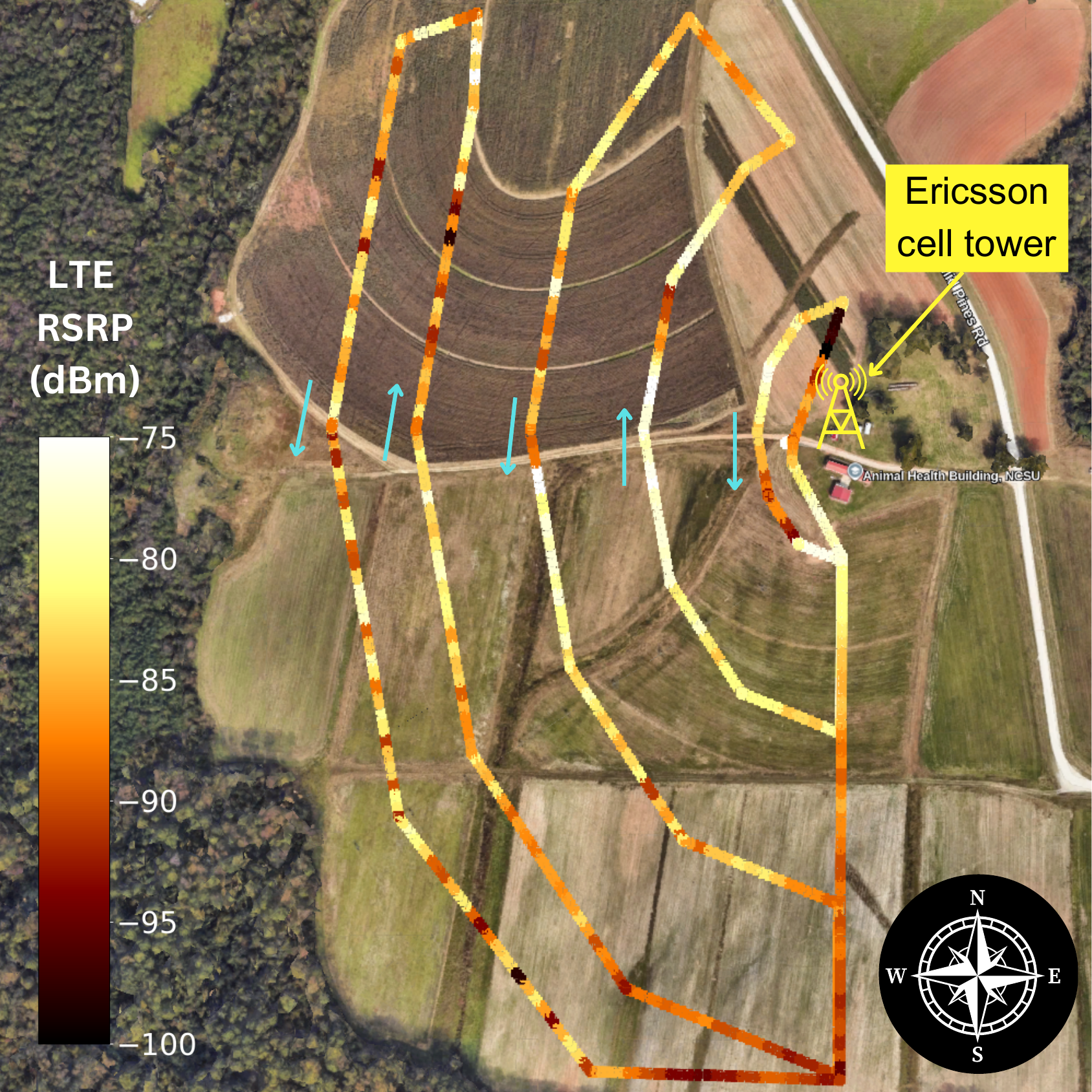}}\hfill
    \subfloat[LTE RSRQ (dB)] {\includegraphics[width=0.33\textwidth]{Images/lte_rsrp_semicircles.pdf}} \hfill
    \subfloat[Cell association]{\includegraphics[width=0.33\textwidth]{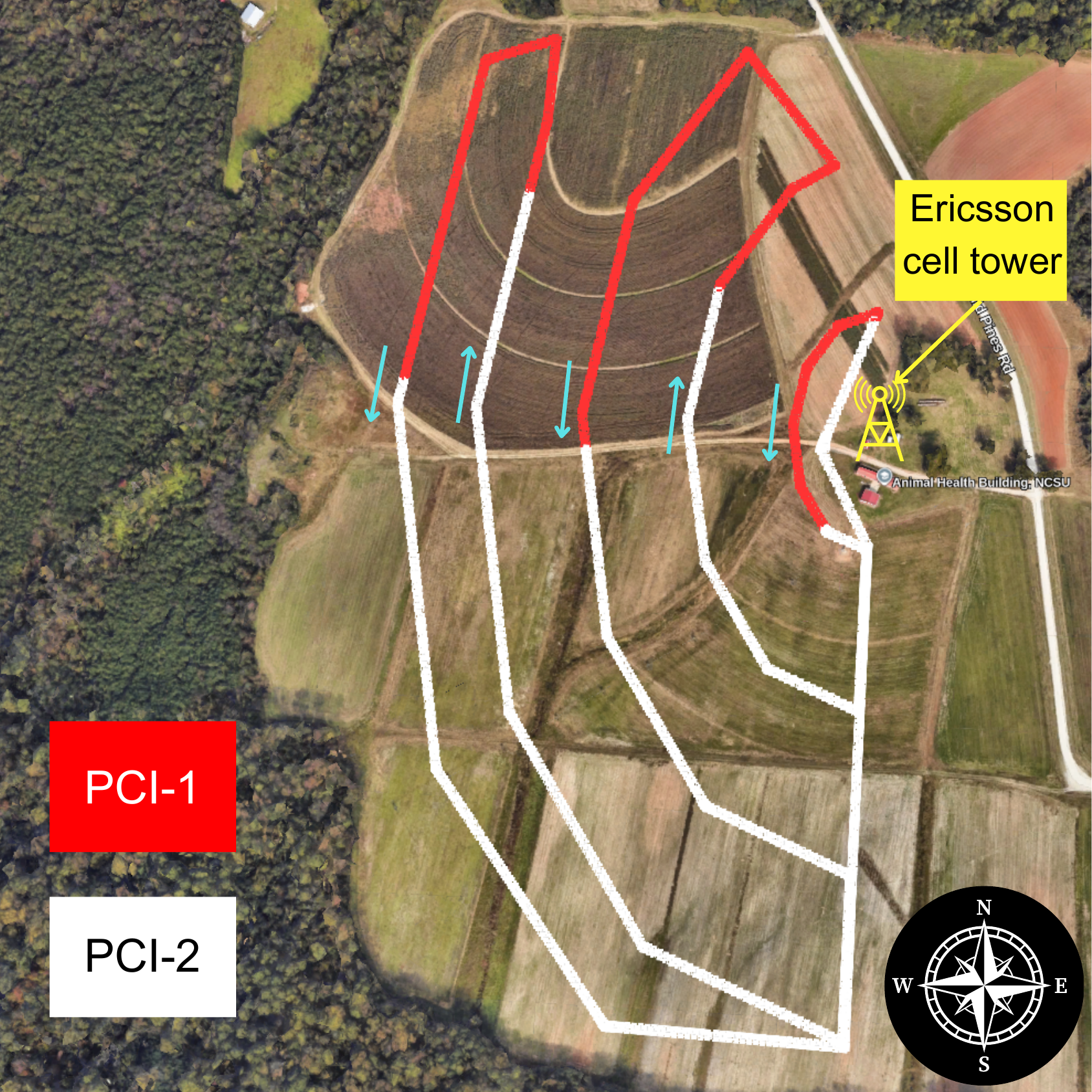}}
      \par \bigskip
    \subfloat[Variation in LTE RSRP with time.] {\includegraphics[width=0.5\textwidth]{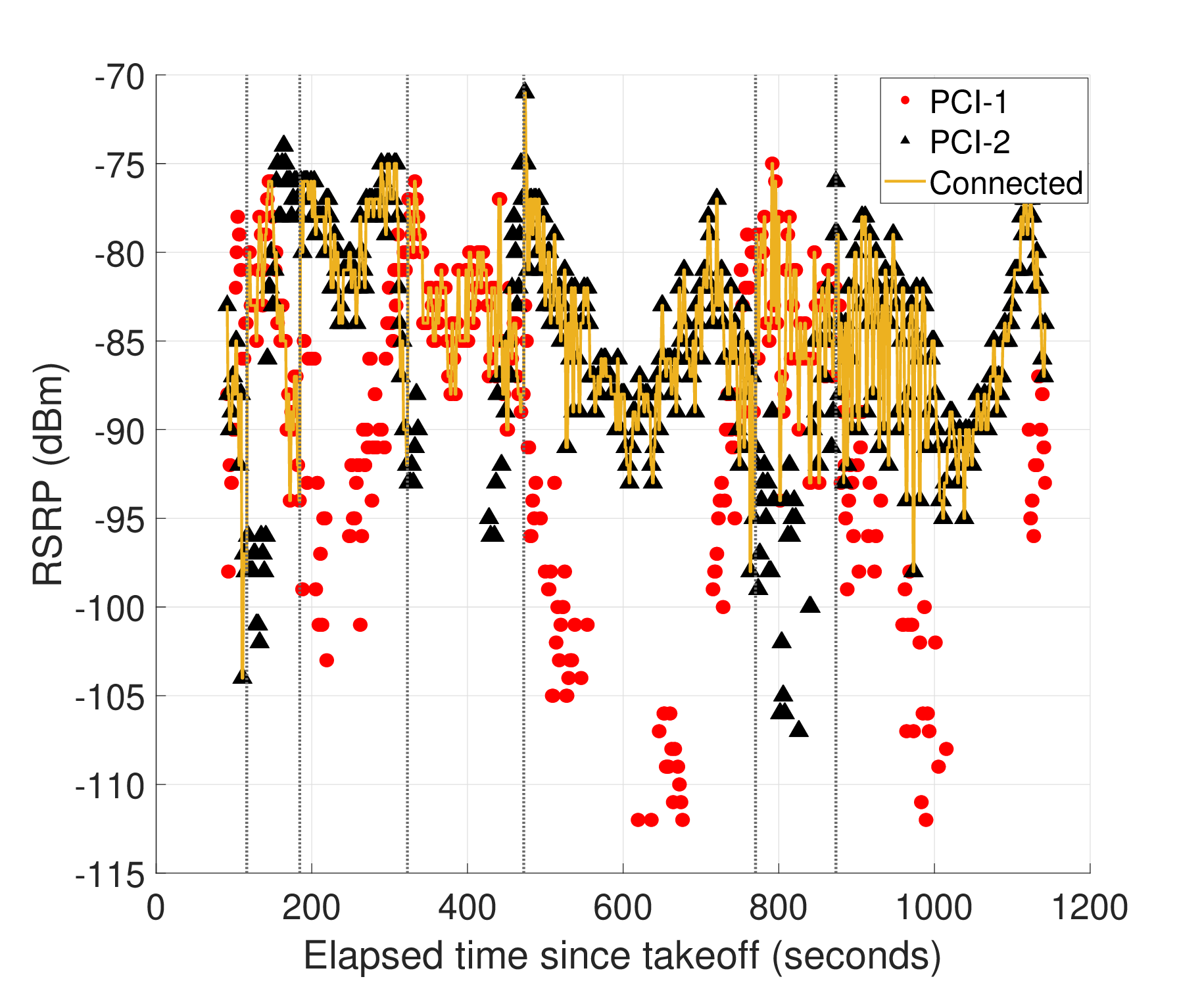}}
    \subfloat[Variation in LTE RSRQ with time, for both cells] {\includegraphics[width=0.49\textwidth]{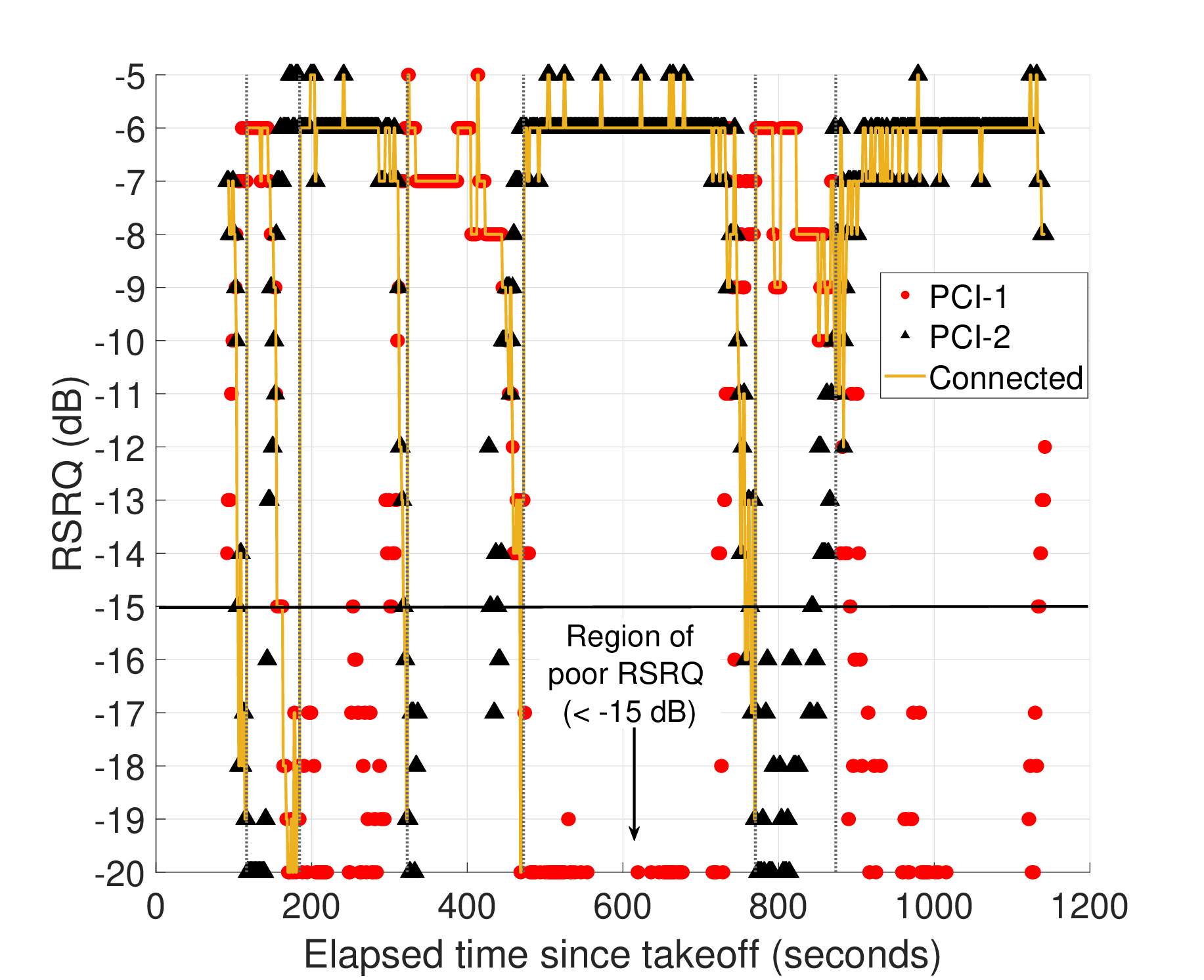}}
    \caption{Heatmaps of LTE RSRP (dBm), RSRQ (dB), and cell associations are shown in (a), (b), and (c) respectively for a polygonal trajectory around the BS, as measured at Samsung S21 device using PawPrints. Variation in LTE RSRP and RSRQ with elapsed time is shown in (d) and (e), respectively, for both LTE cells as well as the cell associated with.}
\label{fig:lake_wheeler_semicircles}
\end{figure*}

The spatial variation in LTE RSRP and RSRQ with the location of the UAV is shown in Fig.~\ref{fig:lake_wheeler_semicircles}, along with the associated cell. Locations associated with PCI-$1$ are shown in red color, and those associated with PCI-$2$ in white. The direction of UAV's travel is shown by blue arrows, to illustrate the impact of the direction of UAV flight on the instant of handover. The temporal variation in LTE RSRP and RSRQ is shown in Fig.~\ref{fig:lake_wheeler_semicircles}d and Fig.~\ref{fig:lake_wheeler_semicircles}e, respectively. RSRP and RSRQ of PCI-$1$ are shown as the red circles, and those of PCI-$2$ as the black triangles. Connected RSRP and RSRQ are shown as the yellow plots. The connected RSRQ falls in the poor region ($< -15$~dB), before handover, and the UAV tends to remain associated with the current cell for a longer duration before handover. 

\begin{figure}[]
    \centering
    \subfloat[5G SS RSRP along horizontal sawtooth trajectory at $30$~m altitude] {\includegraphics[width=0.44\textwidth]{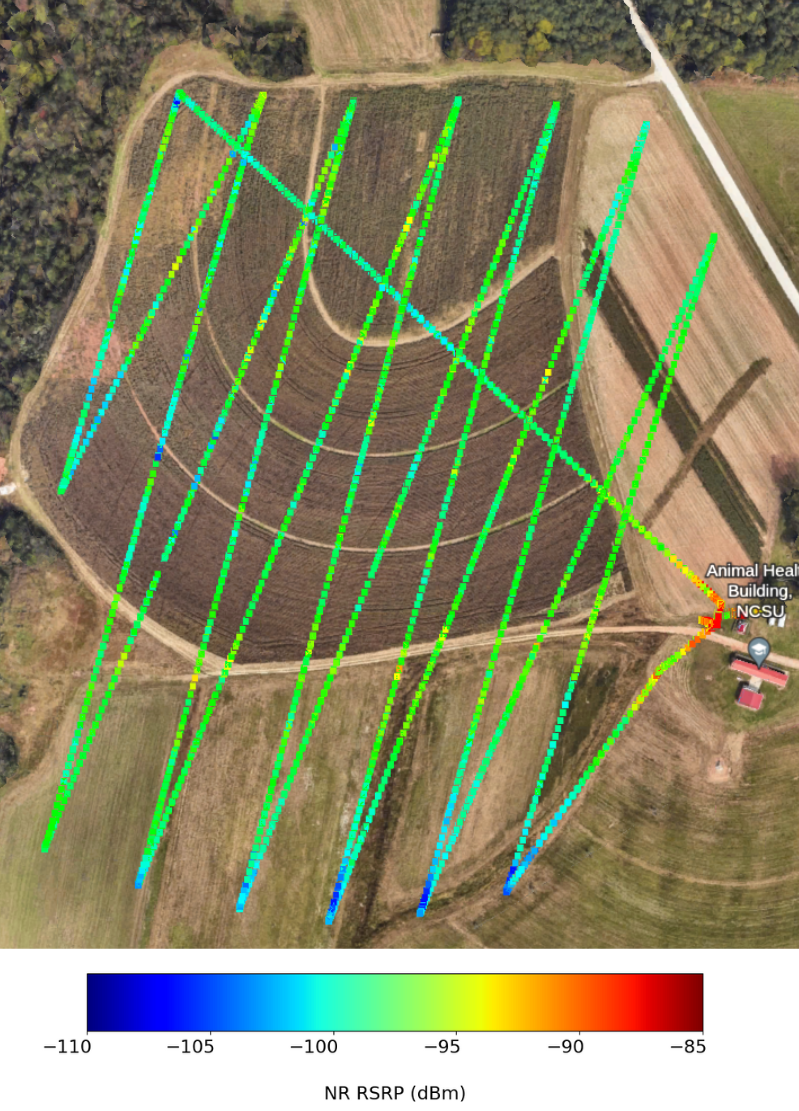}\label{fig:horizontal_sawtooth_rsrp_heatmap_30m}}
    \par
    \subfloat[5G SS RSRP along horizontal sawtooth trajectory at $50$~m altitude] {\includegraphics[width=0.44\textwidth]{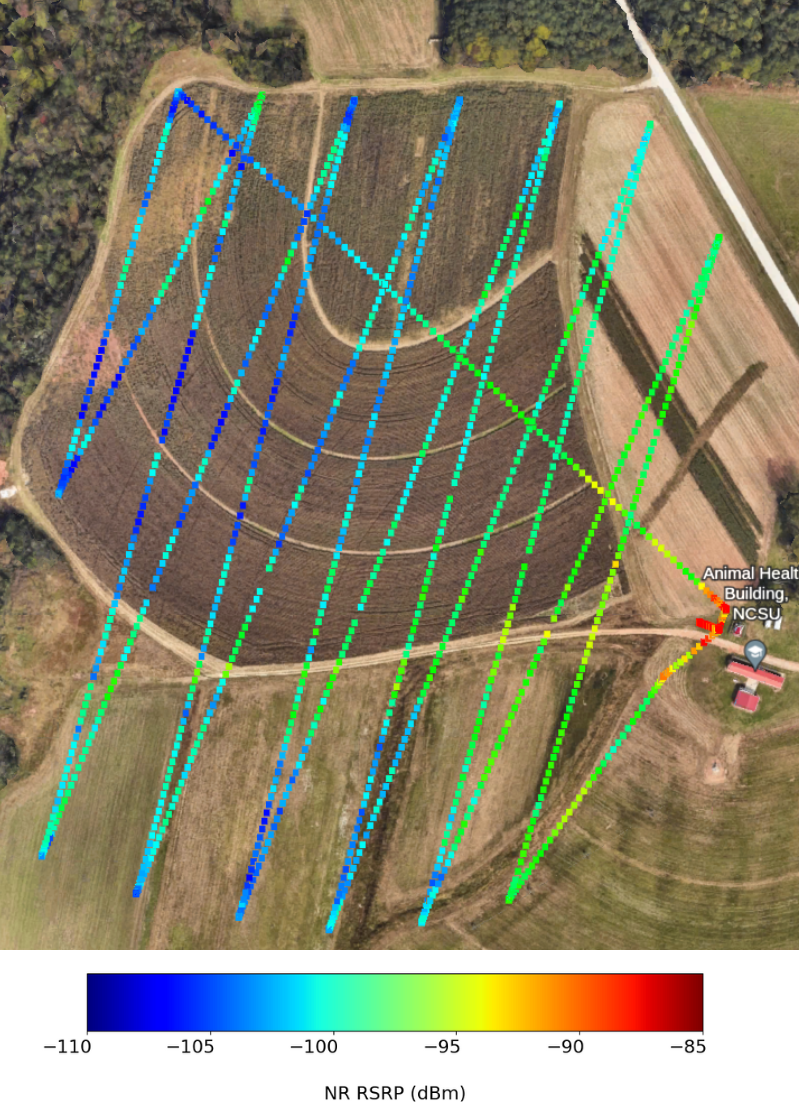}\label{fig:horizontal_sawtooth_rsrp_heatmap_50m}}
    \caption{The heatmap of 5G SS RSRP is shown for the horizontal sawtooth trajectory at an altitude of $30$~m in (a) and at an altitude of $50$~m in (b), as measured at Samsung S23 smartphone using Nemo. }
\label{fig:horizontal_sawtooth_rsrp_heatmap}
\end{figure}

\begin{figure}[]
    \centering
    \subfloat[5G SS SINR along horizontal sawtooth trajectory at $30$~m altitude] {\includegraphics[width=0.44\textwidth]{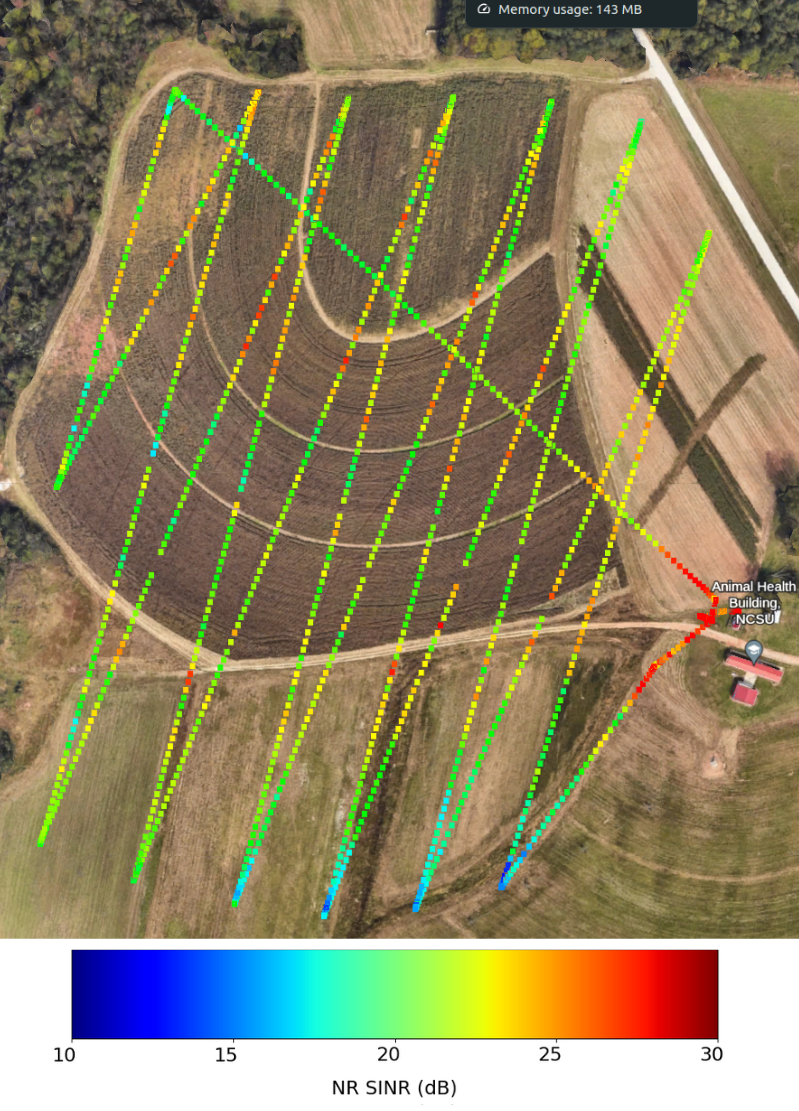}}\label{fig:horizontal_sawtooth_sinr_heatmap_30m}
    \par
    \subfloat[5G SS SINR along horizontal sawtooth trajectory at $50$~m altitude] {\includegraphics[width=0.44\textwidth]{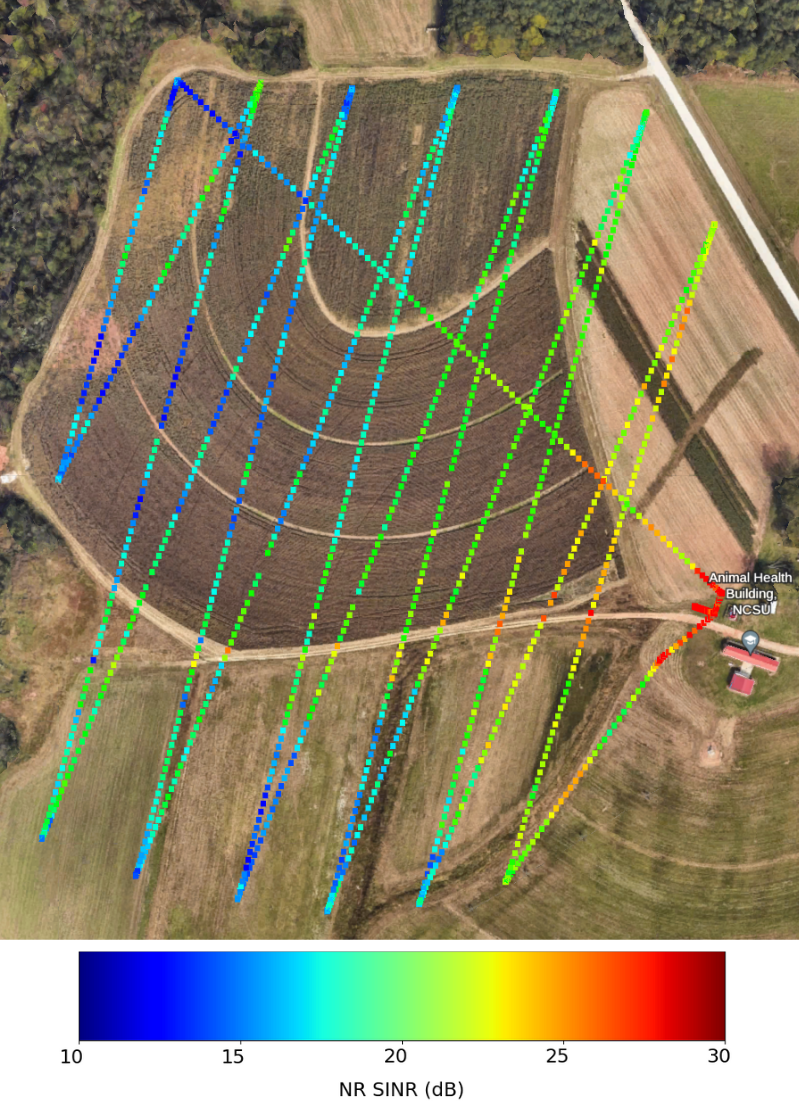}}\label{fig:horizontal_sawtooth_sinr_heatmap_50m}
    \caption{The heatmap of 5G SINR is shown for the horizontal sawtooth trajectory at an altitude of $30$~m in (a) and at an altitude of $50$~m in (b), as measured at Samsung S23 smartphone using Nemo.}
\label{fig:horizontal_sawtooth_sinr_heatmap}
\end{figure}

\begin{figure}[]
    \centering
    \subfloat[5G PDCP throughput along horizontal sawtooth trajectory at $30$~m altitude] {\includegraphics[width=0.44\textwidth]{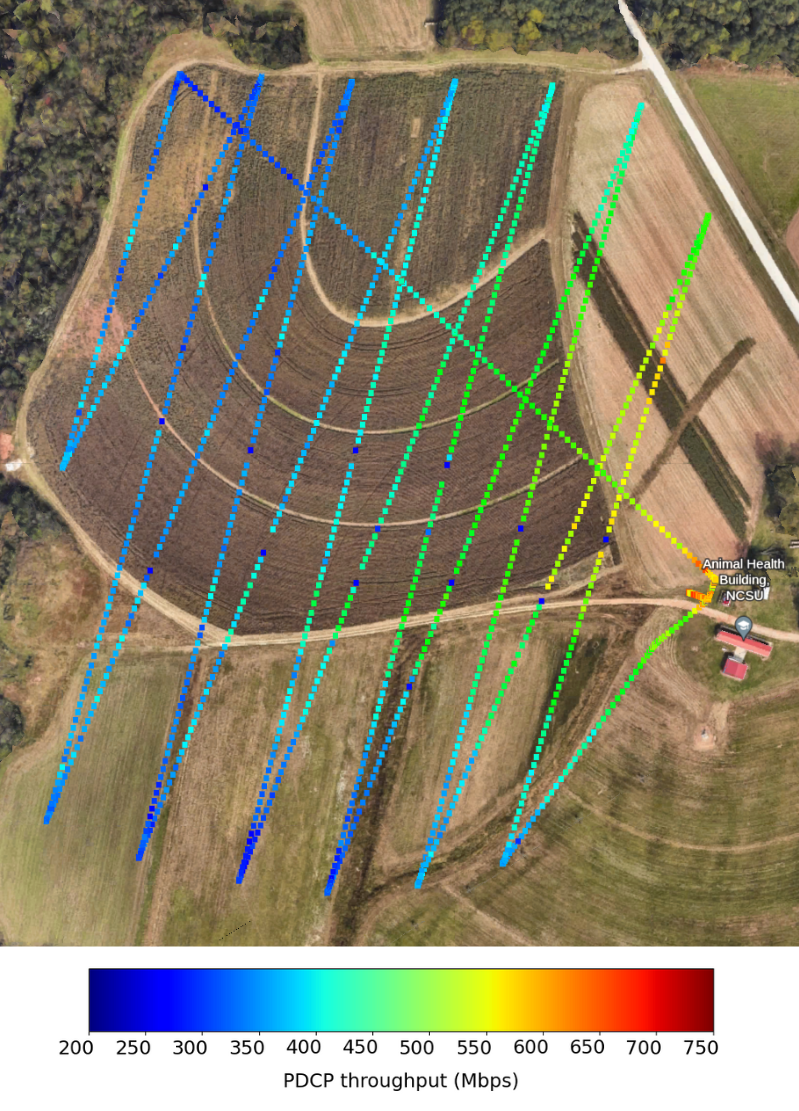}}\label{fig:horizontal_sawtooth_throughput_heatmap_30m}
    \par
    \subfloat[5G PDCP throughput along horizontal sawtooth trajectory at $50$~m altitude] {\includegraphics[width=0.44\textwidth]{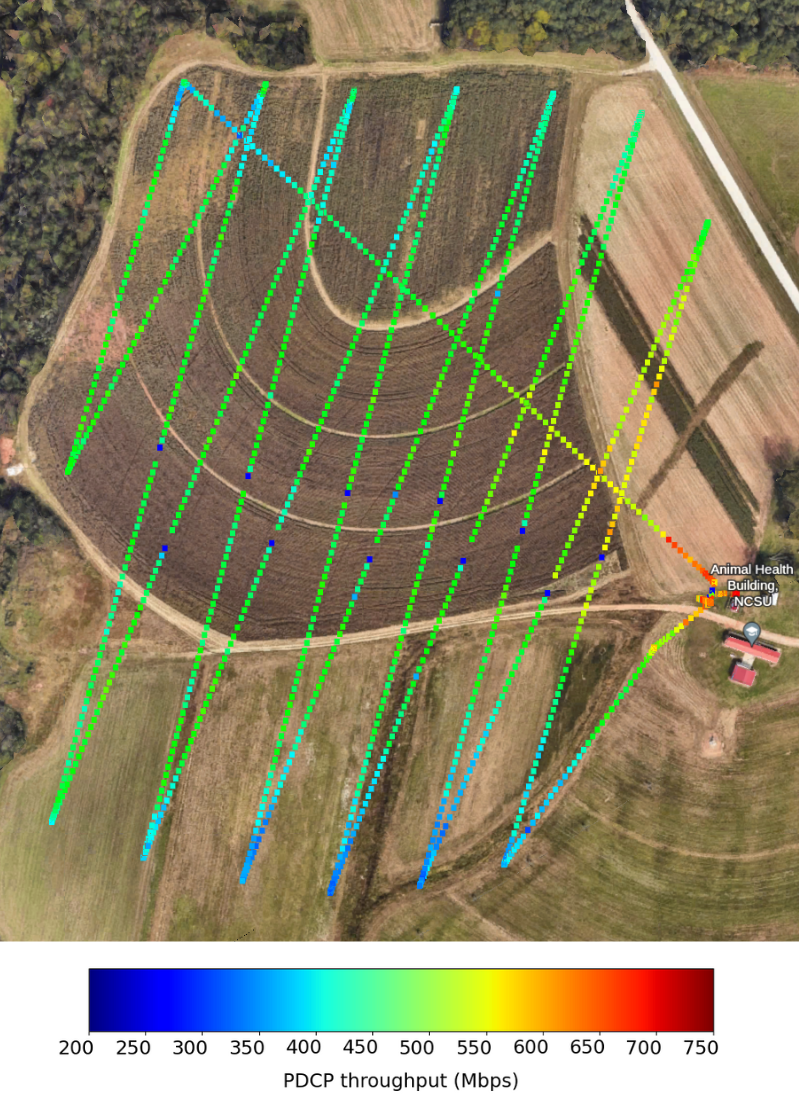}}\label{fig:horizontal_sawtooth_throughput_heatmap_50m}
    \caption{The heatmap of 5G PDCP throughput is shown for the horizontal sawtooth trajectory at an altitude of $30$~m in (a) and at an altitude of $50$~m in (b), as measured at Samsung S23 smartphone using Nemo.}
\label{fig:horizontal_sawtooth_throughput_heatmap}
\end{figure}

\begin{figure}[]
    \centering
    \subfloat[CQI along horizontal sawtooth trajectory at $30$~m altitude] {\includegraphics[width=0.44\textwidth]{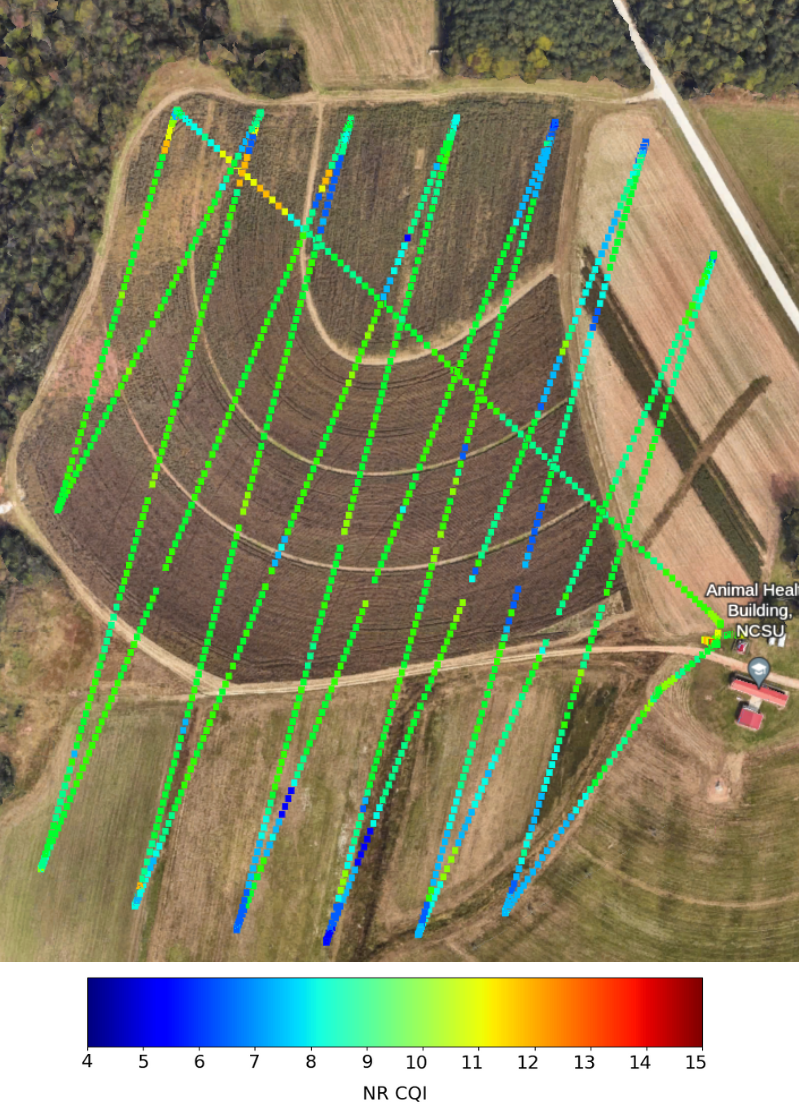}}\label{fig:horizontal_sawtooth_cqi_heatmap_30m}
    \par
    \subfloat[CQI along horizontal sawtooth trajectory at $50$~m altitude] {\includegraphics[width=0.44\textwidth]{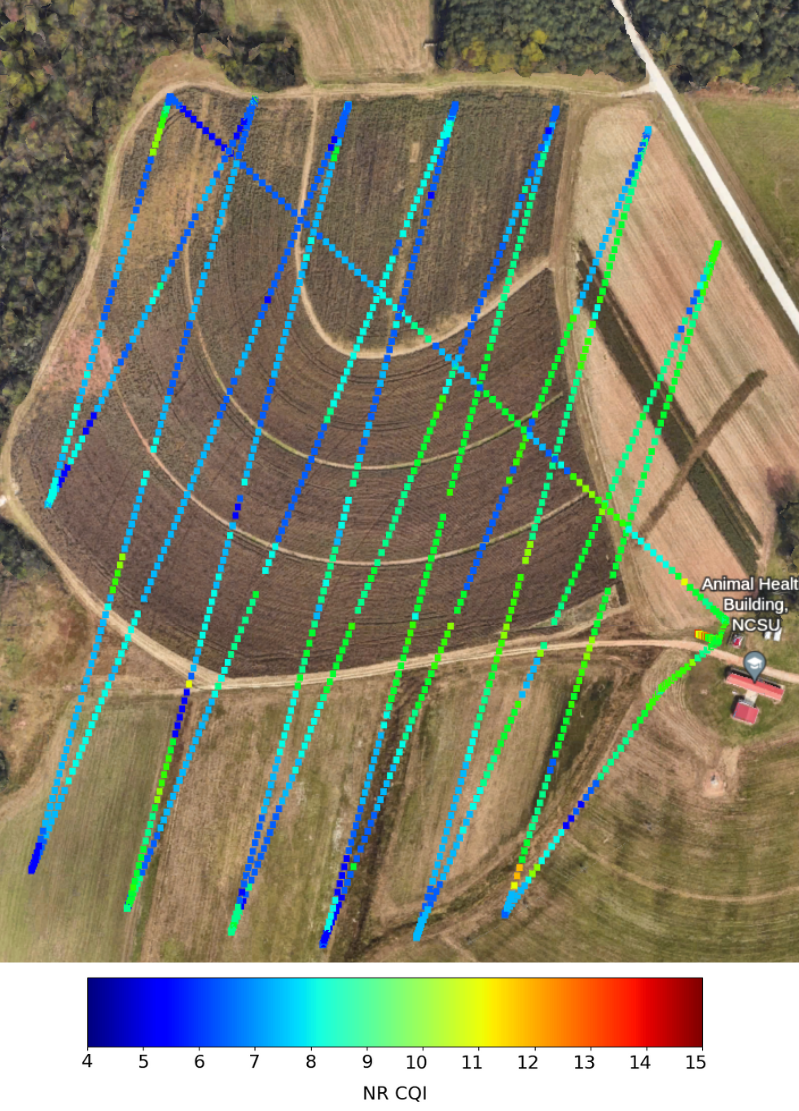}}\label{fig:horizontal_sawtooth_cqi_heatmap_50m}
    \caption{The heatmap of CQI is shown for the horizontal sawtooth trajectory at an altitude of $30$~m in (a) and at an altitude of $50$~m in (b), as measured at Samsung S23 smartphone using Nemo.}
\label{fig:horizontal_sawtooth_cqi_heatmap}
\end{figure}

\begin{figure*}
 \subfloat[PDSCH rank along horizontal sawtooth trajectory at $30$~m altitude] {\includegraphics[width=0.5\textwidth]{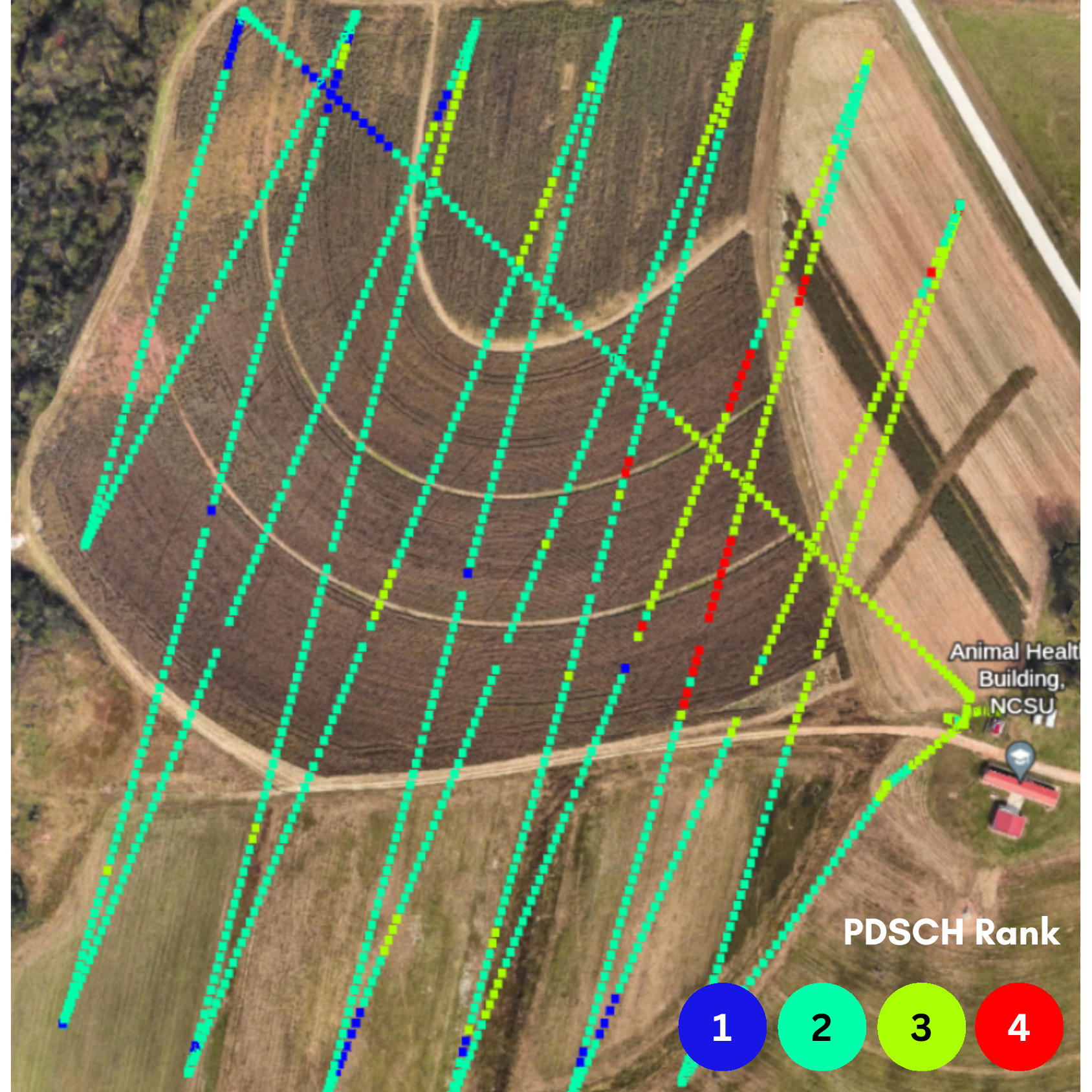}\label{fig:channel_rank_30m}}
    \hfill
    \subfloat[PDSCH rank along horizontal sawtooth trajectory at $50$~m altitude] {\includegraphics[width=0.5\textwidth]{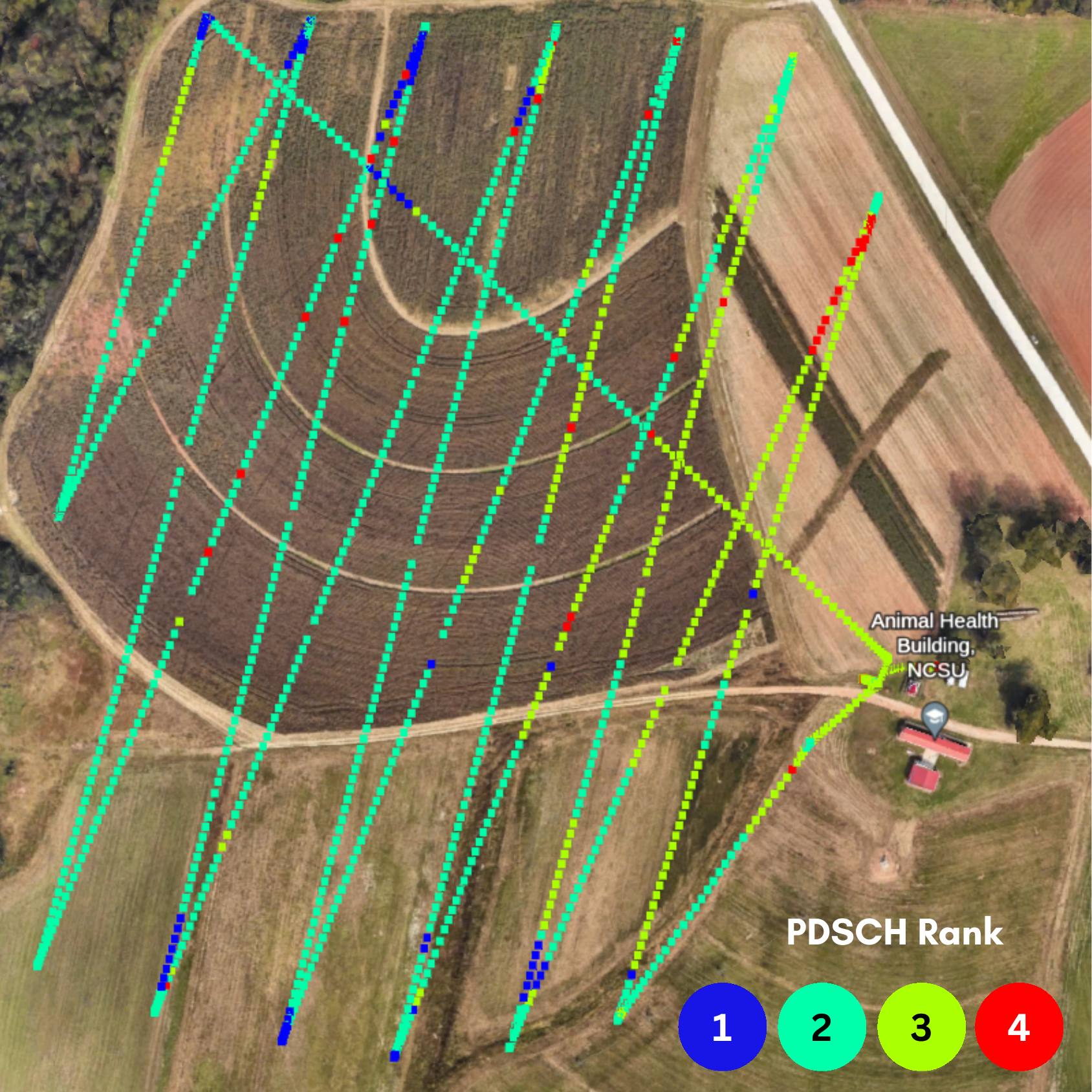}\label{fig:channel_rank_50m}}
    \centering
        \par \bigskip
         \subfloat[Decision plane to distinguish between rank 1 and 4.]
{\includegraphics[width=\textwidth, trim={0cm 0 2.8cm 0},clip]{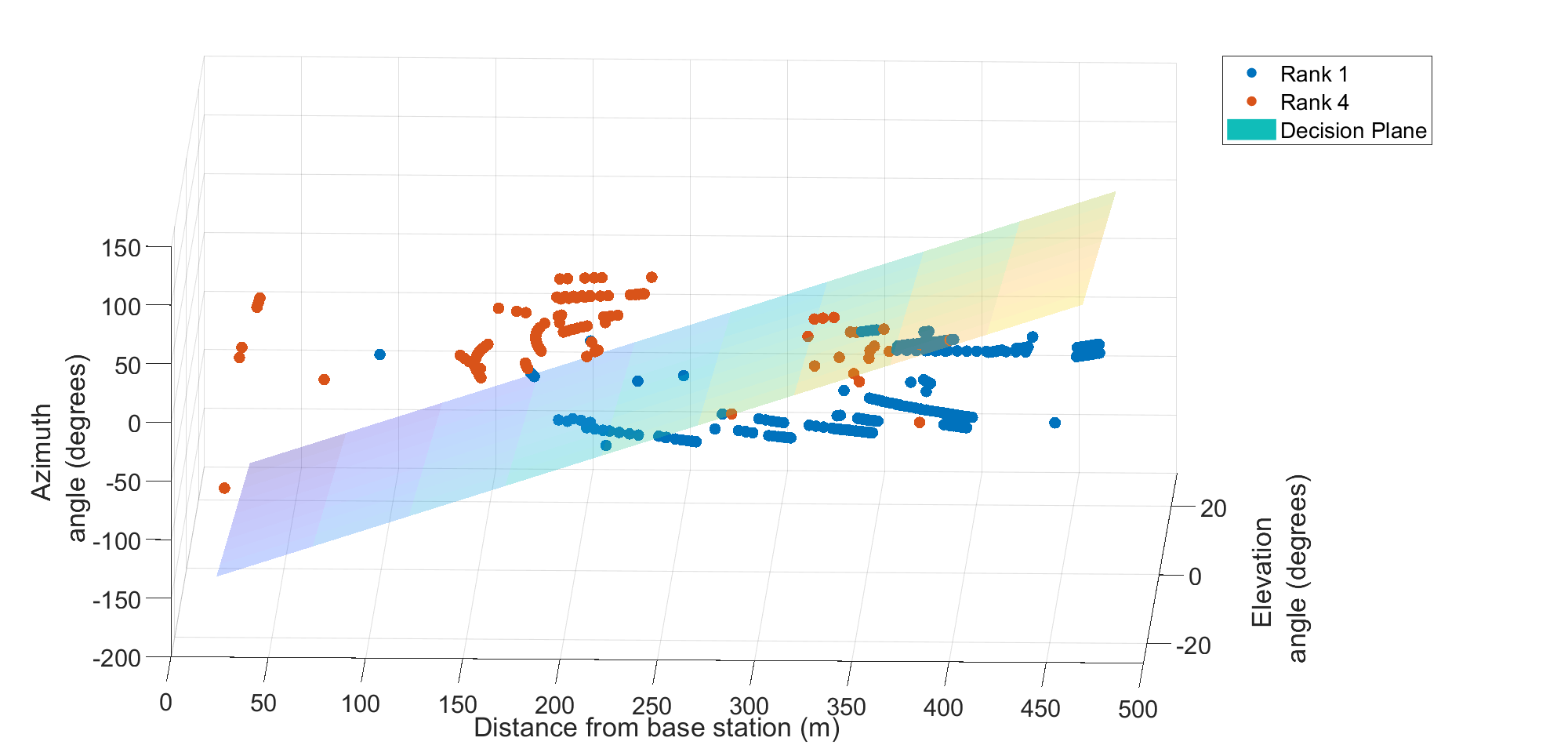}\label{fig:channel_rank_decision_plane}}  
    \caption{The heatmap of PDSCH channel rank is shown for the horizontal sawtooth trajectory at an altitude of $30$~m in (a) and at an altitude of $50$~m in (b), as measured at Samsung S23 smartphone using Nemo. A 3D scatter plot of rank 1 and 4, as a function of the UAV's distance from the BS and angular orientation is shown in (c), along with a decision plane obtained using LDA.}
\label{fig:channel_rank_details}
\end{figure*}

The heatmaps of the variation in 5G RSRP, SINR, packet data convergence protocol (PDCP) throughput, CQI, and channel rank are shown for the horizontal sawtooth trajectory, at altitudes of $30$~m and $50$~m in Figs.~\ref{fig:horizontal_sawtooth_rsrp_heatmap},~\ref{fig:horizontal_sawtooth_sinr_heatmap},~\ref{fig:horizontal_sawtooth_throughput_heatmap},~\ref{fig:horizontal_sawtooth_cqi_heatmap}, and \ref{fig:channel_rank_details}, respectively. A comparison between KPI values at these two altitudes is presented in Table~\ref{tab:kpiComparison}, in terms of the mean difference, standard deviation of difference, and the percentage of flight duration where the KPI measured at $30$~m altitude is greater than at $50$~m. RSRP and SINR at $30$~m are $3.67$~dBm and $3.98$~dB higher than at $50$~m, on an average, and are greater at $30$~m altitude for $77.8\%$ and $79.5\%$ of the flight duration, respectively. The channel rank distribution is similar between the two flights, on an average. The rank was found to be greater at $30$~m in approximately $20\%$ of the time instants, equal in approximately $60\%$ of the instants, and greater at $50$~m altitude for $20\%$. The throughput is also better at $30$~m altitude, by $62.14$~Mbps averaged over all measurement points, and is higher than that at $50$~m for approximately $81\%$ of the flight duration. 

Fig.~\ref{fig:channel_rank_decision_plane} shows a 3D scatter plot of measured channel rank values as a function of distance of the UAV from the BS, and angular orientation. Rank-$1$ is indicated by the blue circles and rank-$4$ by the red circles. A pattern can be seen, where-in locations closer to the BS and near the edge of the main lobe have rank-$4$, and locations farther away from BS but within the main lobe are served with rank-$1$. A decision plane was calculated using linear discriminant analysis, and is shown. This plane resulted in 6\% of the data points being misclassified, out of the total $347$. Five rank-$1$ data points, out of $238$, were misclassified as rank-$4$, while $16$ rank-$4$ data points, out of $109$, were misclassified as rank-$1$. The equation of the LDA plane that separates the rank-$1$ and rank-$4$ measurements, for the considered data set, is:
\begin{align}
 &0.0475 \cdot d_{\mathrm{uav}}(t) -0.1051 \cdot \phi_\mathrm{uav}(t) \\ \nonumber &-0.0892 \cdot \theta_\mathrm{uav}(t) -15.549 = 0.
\end{align} Knowledge about the channel rank distribution can be utilized to enable spatial multiplexing for higher throughput in high-rank conditions or diversity modes to maximize reliability in low-rank scenarios.

\section{Conclusion}\label{section:conclusion}
Aerial measurement campaigns were conducted at the AERPAW testbed using UAVs carrying Android smartphones to measure various signal strength KPIs. The performance of FSPL, polynomial regression, and ML models was compared. Random forest models were found to perform the best, achieving the lowest error scores, of $2.23$~dB RMSE. Additional insights were provided on the variation of cellular KPIs such as 5G RSRP, SINR, PDCP throughput, CQI, and channel rank. Given information about the UAV’s trajectory in
terms of the current position, velocity, heading, and mission
waypoints (if available), future work can predict the signal
quality at the UAV, and make handover decisions intelligently
to minimize severe drops in signal quality. 

\section*{Acknowledgements}\label{sec:acknowledgements}
This work was supported by the NSF PAWR program, under grant number CNS-1939334. The datasets and post-processing scripts to generate the results
in this paper are available at \url{https://aerpaw.org/experiments/datasets/}. The authors would like to thank Thomas Zajkowski, Flight Operations Manager at AERPAW, 
and the entire AERPAW team for their support and contributions.

\begin{table*}[]
\caption{Comparison of KPI values measured during the horizontal sawtooth flights at the two altitudes of $30$~m and $50$~m. $m_{30}$ denotes KPI values measured at an altitude of $30$~m, $m_{50}$ denotes KPI values measured at an altitude of $50$~m.}\def\arraystretch{1.7}
\label{tab:kpiComparison}
\centering
\vspace{-1mm}
\footnotesize
\begin{tabular}{|p{2.2cm}|p{4.3cm}|p{5.2cm}|p{4.2cm} | p{5.0cm} | }
\hline
\textbf{KPI} & \vspace{0.5mm}\textbf{$\overline{m_{30}(t) - m_{50}(t)}$ (mean difference)}  & \textbf{ $\sigma_{m_{30} -m_{50}}$ (standard deviation of difference)} & \textbf{Percentage of flight duration where $m_{30}>m_{50}$}\\
\hline
NR RSRP & 4.69 dBm & 5.11 dBm & 75.96\% \\\hline
NR SINR &  3.5 dB & 4.8 dB &  75.09\% \\\hline
NR CQI  & 1.53  & 2.38  & 66.42 \% \\\hline
PDSCH Rank & 0.04 & 0.61  & 20.07 \% (rank is equal at 62.68\% of the flight duration)\\\hline
PDCP throughput & 62.14 Mbps  & 76.4 Mbps & 81.41 \%
\\
\hline
\end{tabular}
\end{table*}


\vfill \clearpage
\vfill \clearpage
\thebiography
\begin{biographywithpic}
{Simran Singh}{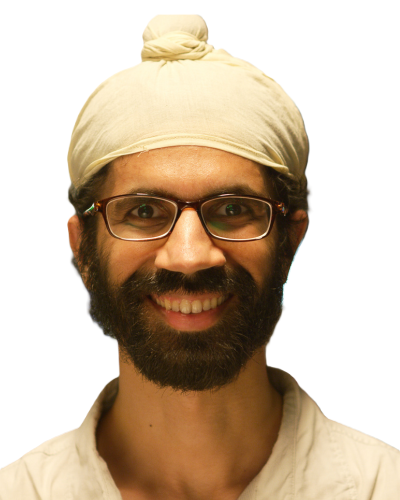} 
\itshape received his B.Tech. degree in Electronics from Veermata Jijabai Technological University Institute, Mumbai, India, in 2011. He worked as a software engineer for four years in the computer-aided design (CAD) domain, building both native and full-stack web applications. He obtained his M.S. in Computer Networks and Ph.D. in Electrical Engineering from NCSU in 2022. He is currently a postdoctoral researcher at NCSU. His research interests include wireless communications for unmanned aerial vehicles.
\end{biographywithpic}

 \begin{biographywithpic}
{An\i l G\"{u}rses}{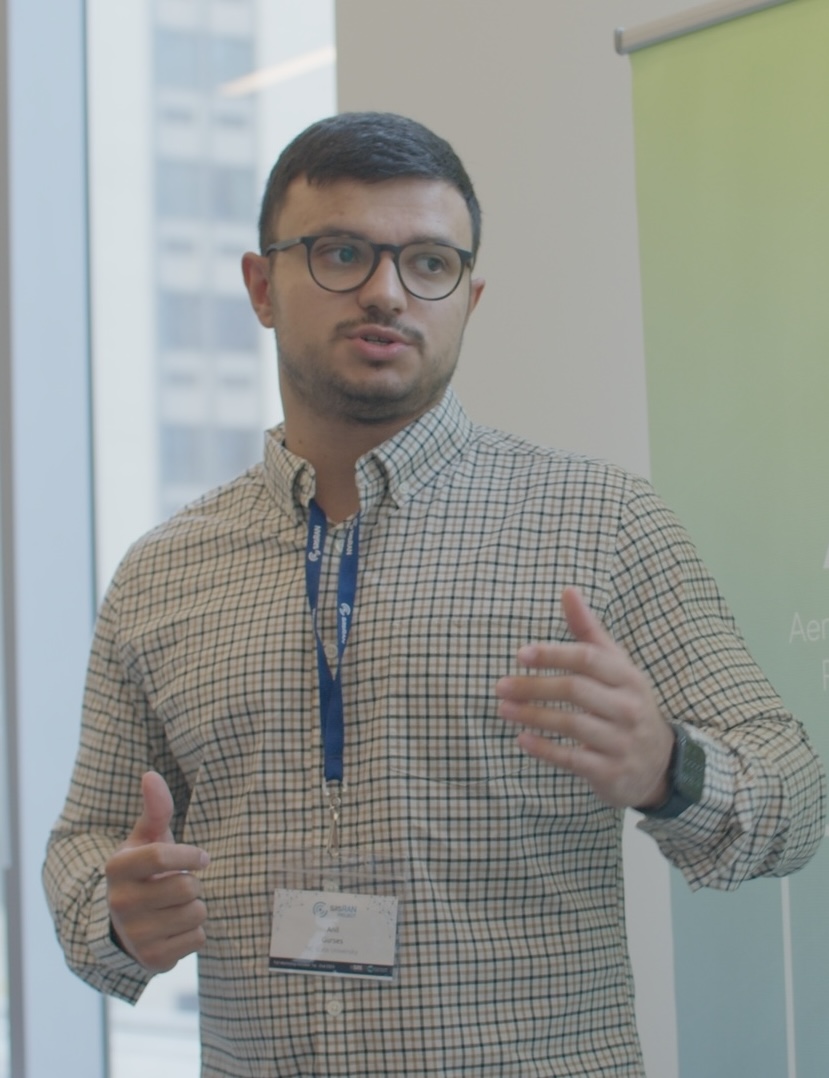} 
\itshape is currently a PhD student in the Electrical and Computer Engineering Department at NC State University, primarily working on UAV communications and Software Defined Radios. One of his key contributions has been developing a digital twin for the AERPAW testbed platform. He holds a bachelor's degree in Electrical \& Electronics Engineering from Istanbul Medeniyet University, during which he also gained industry experience as a software developer at P.I. Works, contributing to 5G Core Network Function development.
\end{biographywithpic}

\begin{biographywithpic}
{Ozgur Ozdemir}{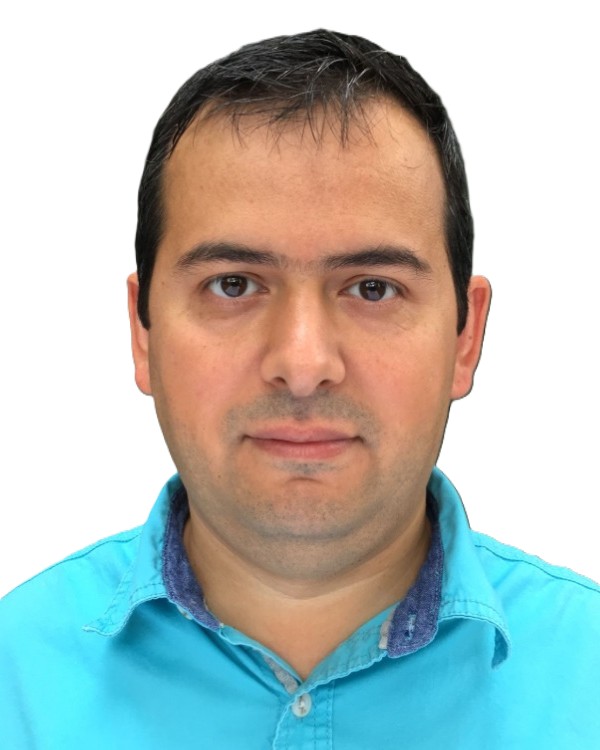}  \itshape
 received the BS degree in Electrical and Electronics Engineering from Bogazici University, Istanbul, Turkey, in 1999 and the MS and Ph.D. degrees in Electrical Engineering from The University of Texas at Dallas, Richardson, TX, USA, in 2002 and 2007, respectively. From 2007 to 2016, he was an Assistant Professor at Fatih University, Turkey, and worked as a Postdoctoral Scholar at Qatar University for 3.5 years. He joined the Department of Electrical and Computer Engineering at NC State as a visiting research scholar in 2017 and is now serving as an Associate Research Professor. His research interests include software-defined radios, channel sounding for mmWave systems, wireless testbeds, digital compensation of radio-frequency impairments, and opportunistic approaches in wireless systems. He is serving as the lead for supporting field experimentation with wireless technologies and drones for the NSF AERPAW platform at NC State.
\end{biographywithpic}

\begin{biographywithpic}
{Mihail L. Sichitiu}{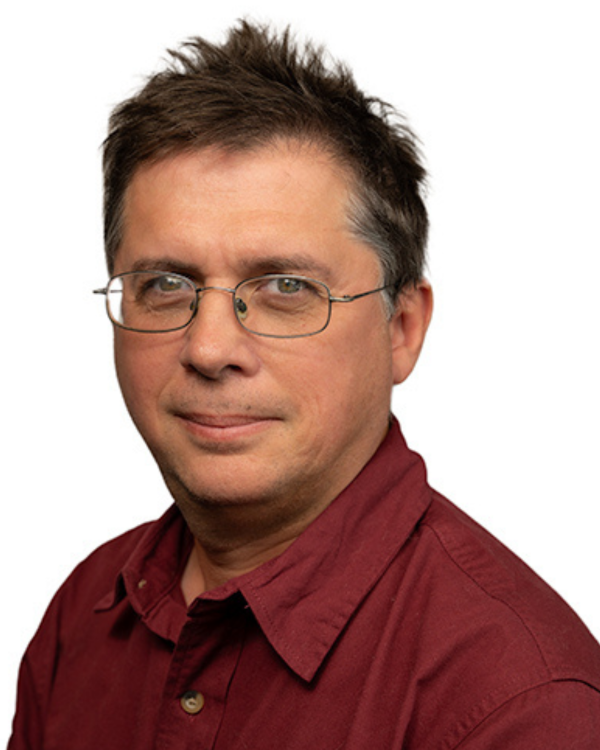} 
\itshape was born in Bucharest, Romania. He received a B.E. and an M.S. in Electrical Engineering from the Polytechnic University of Bucharest in 1995 and 1996 respectively. In May 2001, he received a Ph.D. degree in Electrical Engineering from the University of Notre Dame. He is currently employed as a professor in the Department of Electrical and Computer Engineering at North Carolina State University. His primary research interest is in Wireless Networking with an emphasis on multi-hop networking and wireless local area networks. His recent research interests include UAV networks and digital twins.

\end{biographywithpic}

 \begin{biographywithpic}{\.{I}smail G\"{u}ven\c{c}}{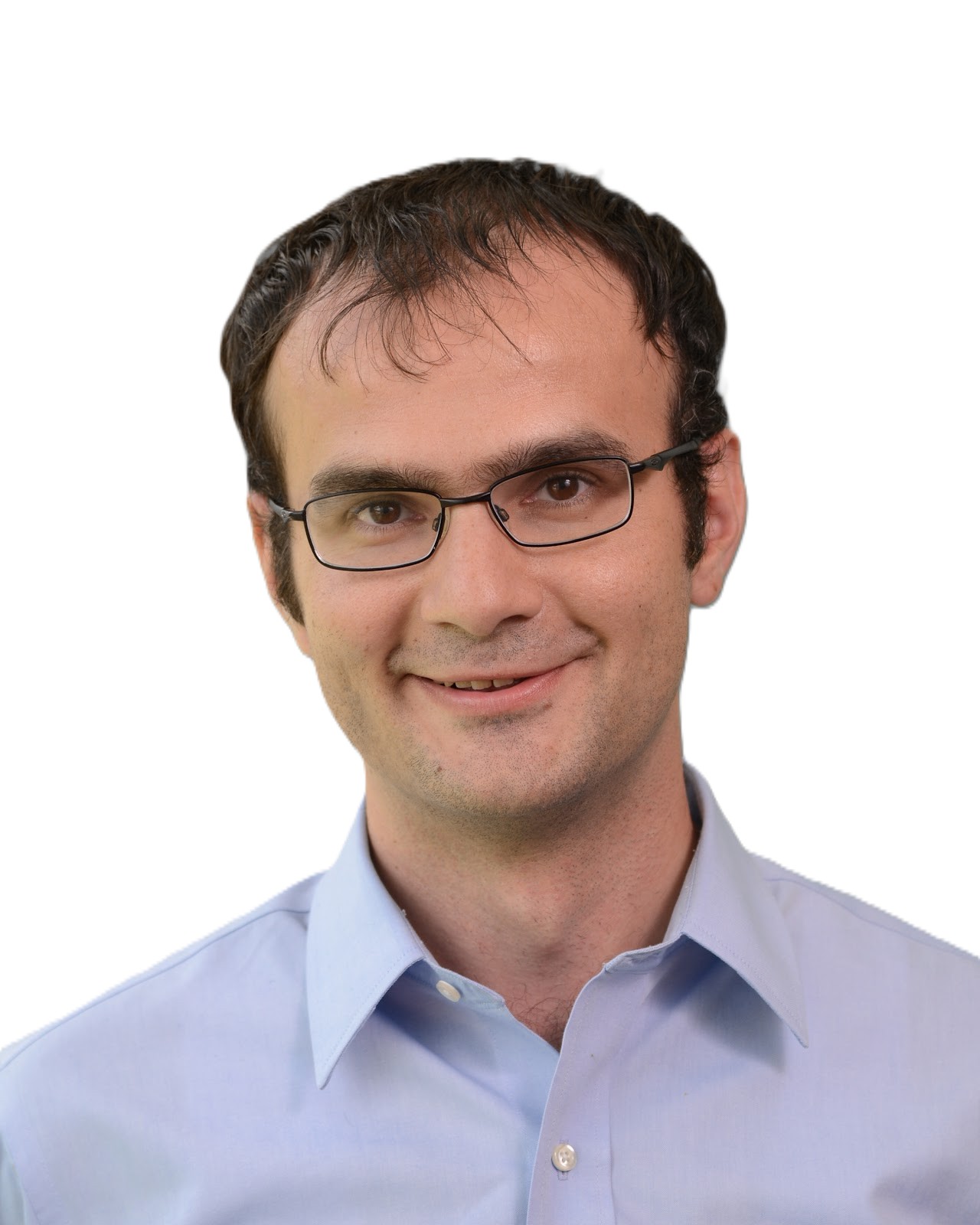}
 \itshape is a Professor at the Department of Electrical and Computer Engineering at NC State University. His recent research interests include 5G/6G wireless networks, UAV communications, millimeter/terahertz communications, and heterogeneous networks. He has published more than 300 conference/journal papers and book chapters, several standardization contributions, four books, and over 30 U.S. patents. Dr. Guvenc is the PI and the director for the NSF AERPAW project and a site director for the NSF BWAC I/UCRC center. He is an IEEE Fellow, a senior member of the National Academy of Inventors, and a recipient of several awards, including NC State University Alcoa Distinguished Engineering Research Award (2023), Faculty Scholar Award (2021), R. Ray Bennett Faculty Fellow Award (2019), FIU COE Faculty Research Award (2016), NSF CAREER Award (2015), Ralph E. Powe Junior Faculty Award (2014), and USF Outstanding Dissertation Award (2006).
\end{biographywithpic}

\begin{biographywithpic}
 {Ram Asokan}
 {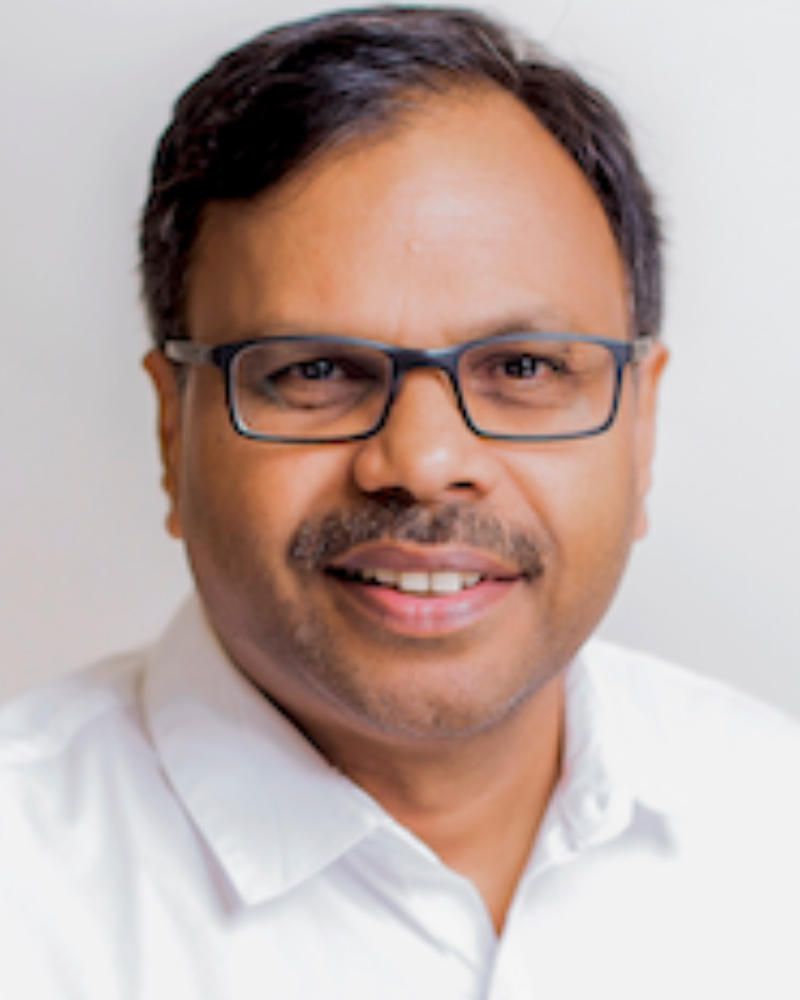}
 \itshape is the Director of Advanced Technology and Senior Architect at Wireless Research Center of North Carolina and a Visiting Scientist at the North Carolina State University. He had worked as Architect at Keysight Technologies, formerly Ixia Communications, from 2008 to to 2017 and previously he was with Ericsson/Sony-Ericsson from 1993 to 2008. Asokan is the inventor/coinventor in some 25 patents and he is a member of Strategic Circle at Asia Open RAN Academy. He is a recipient of Sony Ericsson’s Distinguished Inventor Award, Ixia Engineering Master Award, and Ixia Technical Excellence Award. At Keysight Technologies, Asokan led wireless test system design and development of the 4G LTE and 5G Multi UE sector simulator. At Sony Ericsson and Ericsson, he led product development, Radio access Technology Working Group and 3GPP standardization. He was a key member in architecting the Asia Cellular Satellite (ACeS) GMR-2 geo-mobile satellite system and led the standardization of GMR-2 air interface in ETSI. 
\end{biographywithpic}

\begin{biographywithpic}
 {Rudra Dutta}
 {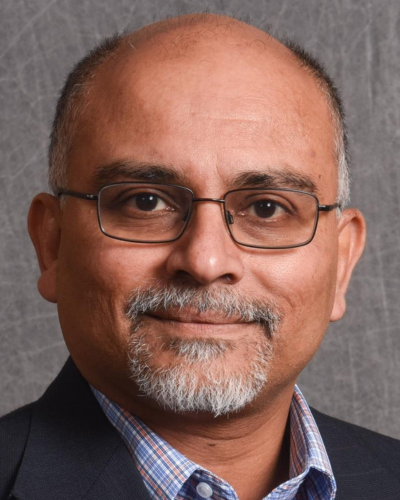}
 \itshape received his Ph.D. in Computer Science from North Carolina State University, Raleigh,
USA, in 2001. From 1993 to 1997 he worked for IBM as a software developer and programmer in various networking related projects. He has been employed as faculty since 2001 in
the department of Computer Science at the North Carolina
State University, Raleigh, since 2013 as Professor, and since
Fall, 2018, serving as Associate Department Head. His current
research interests focus on design and performance optimization of large networking systems, Internet architecture, wireless
networks, and network analytics. He is a senior member of IEEE
and a distinguished member (distinguished engineer) of ACM.
\end{biographywithpic}

 \textbf{Magreth Mushi}
 \itshape is the Program Director of AERPAW. She completed her PhD in Computer Science from NCSU in 2016. She has over a decade of experience leading complex, high-impact projects across communication infrastructure, advanced wireless, digital transformation, and research infrastructure
\end{document}